\documentclass[journal]{IEEEtran}
\usepackage{graphics}
\usepackage{graphicx}
\usepackage{amsfonts}
\usepackage{multirow}
\usepackage{epstopdf}
\usepackage{color}
\usepackage{float}
\usepackage[font=small]{caption}
\usepackage{bm}
\usepackage{algorithm}
\usepackage{algorithmic}
\usepackage{amsmath}
\usepackage{pifont}
\usepackage[utf8]{inputenc}
\usepackage{mathtools}
\usepackage{amssymb}
\usepackage{amsthm}
\usepackage{lipsum}
\usepackage{bbm}
\usepackage{enumitem}   
\usepackage{subcaption}
\usepackage{cite}
\usepackage{steinmetz} 
\usepackage{placeins} 
\usepackage{relsize} 
\usepackage{xcolor, soul}
\sethlcolor{green}
\usepackage{epstopdf}

\usepackage{booktabs}

\theoremstyle{definition}
\newtheorem{definition}{Definition}
\newtheorem{criterion}{Criterion}

\newcommand{\abs}[1]{\left|#1\right|}

\ifCLASSINFOpdf
\else
\fi

\begin{document}

\title{Sparse Convolutional Beamforming for\\ 3D Ultrafast Ultrasound Imaging}

\author{ Regev~Cohen,~\IEEEmembership{Student,~IEEE,} Nitai~Fingerhut, François~Varray, Hervé~Liebgott,
        Yonina~C.~Eldar,~\IEEEmembership{Fellow,~IEEE}
\thanks{Regev Cohen (e-mail:  regev.cohen@gmail.com) and Nitai Fingerhut (e-mail:nitaifingerhut@gmail.com) are with the Department of Electrical Engineering, Technion-Israel Institute of Technology, Haifa 32000, Israel. Fran\c ois Varray and Herv\'e Liebgott are with the Univ. Lyon, INSA-Lyon, Universit\'e Claude Bernard Lyon 1, UJM-Saint Etienne, CNRS, Inserm, CREATIS UMR 5220, U1206, F-69621, Lyon, France. Yonina C. Eldar (e-mail: yonina.eldar@weizmann.ac.il) is with the Faculty of Math and Computer Science,  Weizmann Institute of Science, Rehovot, Israel. This work was supported by the LABEX PRIMES (ANR-10-LABX-0063) and was performed within the frameworks of LABEX CELYA (ANR-10-LABX-0060) of Universit\'e de Lyon, within the program "Investissements d'Avenir" (ANR-11-IDEX-0007) operated by the French National Research Agency (ANR). The RF Verasonics generator was cofounded by the FEDER program, Saint-Etienne Metropole (SME) and Conseil General de la Loire (CG42) within the framework of the SonoCardio-Protection Project leaded by Pr Pierre Croisille. The authors would like to thank LabTAU for their contribution in the development of the 32x32 probe prototype compatible with a driving by 1 to 4 Verasonics Vantage 256 as well as for the provision of two Vantage 256 systems. In addition, this project has received funding from the
European Union’s Horizon 2020 research and innovation program under grant
No. 646804-ERC-COG-BNYQ, and from the Israel Science Foundation under
grant No. 0100101.}}

\maketitle

\begin{abstract}
Real-time three dimensional (3D) ultrasound provides complete visualization of inner body organs and
blood vasculature, which is crucial for diagnosis and
treatment of diverse diseases.    
 However, 3D systems require massive hardware due to the huge number of transducer elements and consequent data size. This increases cost significantly and limits both frame rate and image quality, thus preventing 3D ultrasound from being common practice in clinics worldwide.   
 A recent study proposed a technique, called convolutional beamforming algorithm (COBA), which obtains improved image quality while allowing notable element reduction. COBA was developed and tested for 2D focused imaging using full and sparse arrays. The later was referred to as sparse COBA (SCOBA).
 In this paper, we build upon previous work and introduce a nonlinear beamformer for 3D imaging, called COBA-3D, consisting of 2D spatial convolution of the in-phase and quadrature received signals. The proposed technique considers diverging-wave transmission, thus, achieves improved image resolution and contrast compared with standard delay-and-sum beamforming, while enabling high frame rate. Incorporating 2D sparse arrays into our method creates SCOBA-3D: a sparse beamformer which offers significant element reduction and thus allows to perform 3D imaging with the resources typically available for 2D setups. To create 2D thinned arrays, we present a scalable and systematic way to design 2D fractal sparse arrays. The proposed framework paves the way for affordable ultrafast ultrasound devices that perform high-quality 3D imaging, as demonstrated using phantom and \textit{ex-vivo} data.
\end{abstract}

\begin{IEEEkeywords}
Medical ultrasound, array processing, beamforming, contrast, resolution, sparse arrays, beam pattern, 3D imaging, fractal arrays.
\end{IEEEkeywords}

\section{Introduction}
\label{sec:intro}

\IEEEPARstart{U}{ltrasonography} is a prominent diagnosis technique, commonly used in clinical practices. The low cost and radiation free nature of ultrasound (US) imaging has facilitated its widespread use in medical applications such as obstetrics, cardiology and surgical guidance \cite{fenster2015ultrasound}.

In standard 2D ultrasound imaging, the image is created from multiple scan-lines.
Transducer elements are used to sequentially transmit short acoustic pulses into the medium, focused at different directions. The acoustic signals are reflected back due to tissue perturbations and are received by the array elements. Upon reception, the signals are sampled and digitally beamformed to yield a line in the image. This process is repeated for consecutive directions to create the complete image frame.     

The performance of the above approach, used by most commercial US scanners, is characterized by the following major aspects: image quality (i.e. resolution and contrast), frame (or volume) rate and processing rate. The common beamfomer, delay-and-sum (DAS) \cite{thomenius1996evolution,karaman1995synthetic}, is widely used due to its simplicity and real-time capabilities, but it suffers from poor resolution and contrast. The number of transmit-receive sequences required to build all scan-lines is typically several hundreds in 2D settings. This limits the frame rate to tens of frames per second, making it insufficient for cardiac applications such as the proper evaluation of the fastest phenomena in the cardiovascular system (flow patterns in the aorta or pulse wave propagation), or shear wave elastography. A common approach to improve frame rate is the use of ultrafast imaging. Here, several tilted plane-waves or diverging-waves (DWs) are sequentially transmitted. Upon reception, a beamformed signal is created by DAS after each transmission, and the signals are then summed coherently to yield a final compounded image. This leads to a dramatic increase in frame rate while providing improved image resolution and contrast. However, since the entire region of interest is reconstructed following each transmission, this strategy increases the processing rate and exhibits a large computational load which typically requires the use of graphics processing units \cite{eklund2013medical}.         

Conventional 2D US is highly operator-dependent as it relies on the physician's knowledge of the human anatomy and her or his expertise to comprehend 3D anatomic structures from several planar 2D images. Performing 3D imaging reduces operator dependence since once the volumetric data is obtained, any arbitrary view of the data can be displayed, including anatomical structures within it that are intrinsically 3D. However, 3D ultrasound necessitates the use of 2D probes where the number of elements exceeds several thousands. The latter implies a massive increase in data size and processing rates which may degrade frame rate and image quality. Furthermore, current 3D imaging requires cumbersome hardware that is only available in a few research facilities (e.g. the parallelized Verasonics systems at the University of Lyon \cite{petrusca2018fast}, the SARUS scanner at the Technical University of Denmark in Lyngby \cite{jensen2013sarus}, the parallelized Aixplorer systems at the Langevin Institute in Paris \cite{provost20153}). 

Due to the mentioned limitations, it is of utmost importance to develop efficient methods to perform high frame-rate 3D imaging with limited hardware. This will enable the daily use of 3D imaging in clinics worldwide.


Several techniques have been proposed that aim at reducing the large amount of receive channels. Savord and Solomon proposed a strategy called microbeamforming \cite{savord2003fully, wildes20164, santos2016diverging, matrone2014volumetric, bhuyan2013integrated, kortbek2013sequential, fisher2005reconfigurable, wygant2009integrated, smith1991high, von1991high, lok2018microbeamforming}, where the array elements are divided to sub-arrays which are analog-beamformed. However, the latter requires custom expensive integrated circuits that exhibit high power consumption \cite{savord2003fully,fuller2005sonic,lee2004miniaturized}. Moreover, the acquisition flexibility is reduced due to the predetermined delays associated with the sub-arrays.
In \cite{chen2011cmut}, the authors proposed the use of 2D row-column-addressed arrays  \cite{chen2011cmut,rasmussen20133d,rasmussen20133,rasmussen20153,christiansen20153, daya2017compensated, bouzari2017curvilinear, flesch20174d, logan201132, savoia2007p2b}, in which every row and column in the array acts as one large element. However, large elements may exhibit significant edge effects that limit image quality \cite{rasmussen20133d}. The notion of separable beamforming was introduced in \cite{yang2014separable} and \cite{owen2012application} wherein 2D beamforming is performed by two separable 1D steps which facilitates the computation but the overall amount of data remains the same.    

Another approach, adopted from sonar processing, is synthetic aperture \cite{nikolov2003investigation, jensen2006synthetic, kortbek2008synthetic, wygant2006beamforming, savord2003fully} (SA) which performs channel multiplexing to address a full 2D array with a small number of electronic channels. In this context, a method called multi-element synthetic
transmit aperture (MSTA) was introduced in \cite{lokesh2019diverging} where
unfocused or diverging-waves are transmitted using a limited number of active elements, while all the elements are utilized upon reception. In \cite{bottenus2013synthetic}, SA was combined with short-lag spatial coherence. However, these techniques use all array elements on reception.

A promising framework for data reduction is compressed sensing (CS) \cite{eldar2012compressed,eldar2015sampling} which includes the concept of analog Xampling \cite{tur2011innovation, zhuang2012ultrasonic, zhou2013compressed, liebgott2013pre, achim2010compressive, tzagkarakis2013joint, quinsac2012frequency, wagner2012compressed, chernyakova2014fourier, gedalyahu2011multichannel, baransky2014sub}. Such techniques focus on reducing the sampling rate by assuming the ultrasound signal can be sparsely represented in some chosen basis. The reconstruction performance of ultrasound signals in different bases was invesitgated in \cite{liebgott2013pre}. A method for reducing the sampling rate was developed in \cite{wagner2012compressed} where the authors described the ultrasound echoes within the finite rate of innovation framework as a small number of replicas of a transmitted pulse \cite{eldar2012compressed,eldar2015sampling}. This approach was exploited to develop sub-Nyquist data acquisition \cite{chernyakova2014fourier}, including plane-wave imaging \cite{chernyakova2018fourier} and volumetric  imaging \cite{burshtein2016sub}. A different beamforming method, called compressed sensing based synthetic transmit aperture \cite{liu2017compressed}, consists of transmitting a small number of randomly weighted plane-waves, thus increasing frame-rate, and using CS techniques for recovering  the full channel data. Yet, none of the above considered  receive element reduction. 

 An alternative interesting strategy is performing DAS beamforming with sparse arrays, where some of the elements are removed, including both random arrays and deterministic designs \cite{davidsen1994two,brunke1997broad,yen2000sparse,austeng2002sparse,karaman2009minimally,diarra2013design,ramadas2014application,emmanuel2017validation,mitra2010general}. However, implementing DAS with random thinned arrays typically leads to increased average side lobe levels. Given a desired number of active elements, 2D sparse arrays can be optimized \cite{roux2016wideband, roux20162, diarra2013design, tekes2011optimizing, karaman2009minimally} to produce homogeneous imaging capability over the entire volume of interest. Still, such sparse arrays exhibit lower sensitivity compared to full arrays \cite{roux2018experimental}. In addition, the array design is typically not scalable and has to be repeated for each setting.    


Following the line of works on sparse arrays \cite{cohen2018sparse, cohen2018optimized, cohen2017sparse,  davidsen1994two,brunke1997broad,yen2000sparse,austeng2002sparse,karaman2009minimally,diarra2013design,ramadas2014application,emmanuel2017validation,mitra2010general}, Cohen \textit{et. al.} \cite{cohen2018coba} introduced a convolutional beamforming algorithm (COBA) based on the convolution of the delayed RF signals which can be implemented at 
low-complexity using the fast Fourier transform (FFT). COBA creates a virtual array, termed the sum co-array \cite{cohen2018optimized}, which dictates the beamforming performance. For an appropriate element-arrangement, the resultant sum co-array may be larger than the physical array, leading to a notable improvement in image resolution and contrast, compared to standard DAS. Based on this, a sparse version of COBA can be used with a small number of elements, referred to as sparse COBA (SCOBA). SCOBA achieves a significant decrease in the number of elements without compromising image quality. COBA and SCOBA have been implemented in the context of 2D imaging with focused transmission, hence, they exhibit low frame rate and operator dependency characteristic of all 2D methods. 

In this work, we extend the notion of convolutional beamforming to 3D imaging with diverging-wave transmission to allow ultrafast frame-rate. We introduce a non-linear beamformer, referred to as COBA-3D, which performs coherent compounding upon reception and then computes the 2D spatial convolution of the resultant in-phase and quadrature (IQ) signals using 2D FFT. We show that COBA-3D achieves improved 3D image resolution and contrast in comparison to DAS. Incorporating 2D sparse arrays into our framework leads to sparse COBA-3D (SCOBA-3D) which in turn provides significant element reduction, allowing to preform high-quality 3D imaging with the resources typically available in 2D settings. Our approach relies on the design of 2D sparse arrays. To address this challenge, we present a simple recursive scheme for constructing arbitrarily large 2D fractal arrays \cite{cohen2020sparse, cohen2019sparse, puente1996fractal, werner1999fractal,werner2003overview,feder2013fractals,falconer2004fractal} on which SCOBA-3D can operate.

We validate the proposed methods using phantom scans which include point-reflectors and an anechoic cysts to assess image resolution and contrast. We show qualitatively and quantitatively that COBA-3D outperforms standard DAS with coherent compounding. Results obtained by SCOBA-3D prove that we can utilize an order-of-magnitude lower number of receive elements, reduced from 961 elements composing the full array to just 169 ($\approx18\%$), without compromising image quality. To strengthen our results we show images obtained from \textit{ex vivo} data, setting the path towards real-time clinical application of the proposed methods.       

The rest of the paper is organized as follows. In Section\,\ref{sec:model}, we derive an expression for the 2D beam pattern and formulate our problem. Section \ref{sec:3Dcoba} describes the convolutional beamformer for 3D imaging with diverging-waves transmission. We further describe the use of sparse arrays to obtain element-reduction and present our fractal array design. In Section \ref{sec:results}, we evaluate  the performance of our beamformers in different settings using phantom and \textit{ex-vivo} data. Finally, Section \ref{sec:conclude} concludes the paper. 

\section{Beam Pattern and Problem Description}
\label{sec:model}

\subsection{Beam Pattern}
\label{subsec:model}

We begin by presenting the concept of a 2D US beam pattern which provides a mean for the design and evaluation of beamformers and the transducer arrays on which they operate. The presentation is based on \cite{cohen2018coba}, extended to the 3D setting. 

Consider a 2D uniform planar array (UPA) whose sensors are located in the $xy$ plane at 
\begin{equation}
p_{n,m}=(nd_x,md_y,0),\quad n\in [-N, N],\, m\in[-M, M],
\end{equation}
where $N,M\in\mathbb{N}^+$, $d_x$ and $d_y$ are the element spacing (pitch) in the $x$ and $y$ directions respectively, and $z$ represents the axial axis. 

At reception, consider a scatterer located at $(r,\theta, \phi)$ where $r$ is the distance from the center of the array, $\theta$ and $\phi$ are the azimuth and elevation angles respectively. An acoustic pulse is reflected off the scatterer and propagates back through the tissue at the speed of sound $c$, assumed to be constant.
The backscattered signal is received by the transducer elements where the time of arrival at each sensor depends on the position of the scatterer and the array geometry. The signal received by the centeric element is denoted by
\begin{equation}
f(t)=h(t)e^{j\frac{2\pi}{\lambda}ct},
\end{equation}  
where $h(t)$ is the signal envelope, and $\lambda$ is the transducer wavelength. Assuming the scatter is at the array far-field and the envelope is narrow-band \cite{cohen2018coba}, we can express the signal, received by the element positioned at $p_{n,m}$, as   

\begin{equation}
f_{n,m}(t)=h(t)e^{j\frac{2\pi}{\lambda}c(t-\tau_{n,m})}=f(t)e^{-j\frac{2\pi}{\lambda}c\tau_{n,m}},
\label{eq:received signal}
\end{equation}
where the time delay is given by
\begin{equation}
   \tau_{n,m}=\frac{\sin\theta\left(nd_x\cos{\phi}+md_y\sin{\phi}\right)}{c}. 
   \label{eq:delay}
\end{equation}

Beamforming is the process in which the received signals are temporally filtered and then combined to create the final image. The conventional beamformer is DAS which applies appropriate delays on each of the received signals according to a certain direction of interest $(\theta_0, \phi_0)$, and then performs a weighted sum of the results. This creates a beamformed signal given by
\begin{equation}
y(t)=\sum_{n=-N}^{N}\sum_{m=-M}^{M} w_R[n,m]f_{n,m}(t+\gamma_{n,m}),
\label{eq:beamformedsignal}
\end{equation}
where $w_R[n,m]$ are the weights (upon reception) and the time delays are given by 
\begin{equation}
\gamma_{n,m}= \frac{\sin\theta_0\left(nd_x\cos{\phi_0}+md_y\sin{\phi_0}\right)}{c}.
\end{equation}
Collecting the beamformed signals of all desired directions allows to construct the entire 3D image and display any required view of the scanned volume.   

\begin{figure}[h]
 \centering
 \includegraphics[trim={2cm 3cm 3cm 5cm},clip,height = 4cm, width = 0.9\linewidth]{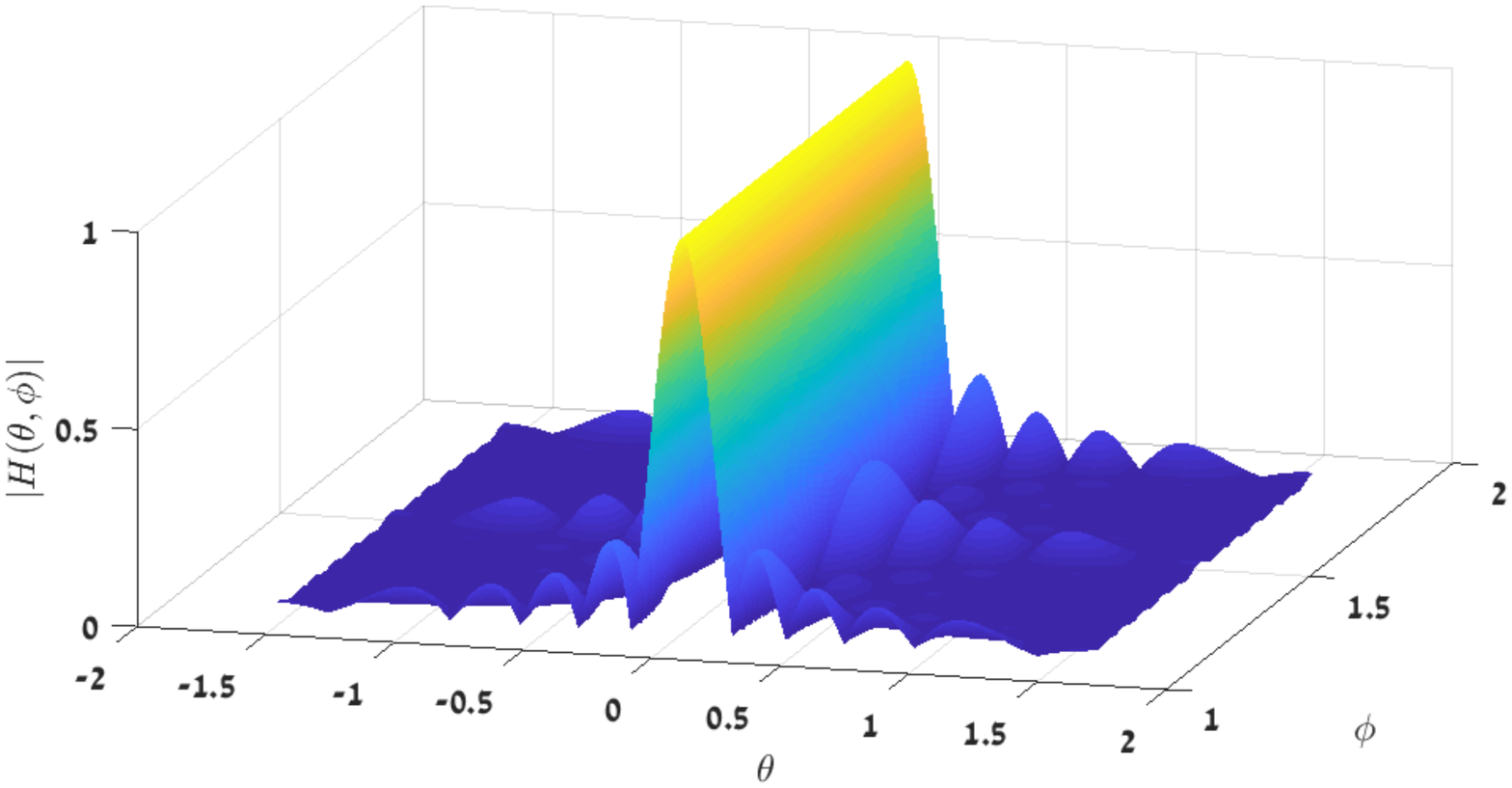}
 \caption{Magnitude of an example beam pattern generated by DAS with unity receive weights.}
  \label{fig:beampattern}
 \end{figure}

Assuming the input signal is a unity amplitude plane wave $f(t)=e^{j\omega_0t}$ where $\omega_0=\frac{2\pi}{\lambda}c$, the expression for the receive beam pattern is given by \cite{fenster2015ultrasound}  

\begin{align}
\begin{split}
H_{RX}(\theta,\phi)&\triangleq\sum_{n,m} w_R[n,m]e^{-j\tau_{n,m}} \\ &=
\sum_{n,m} w_R[n,m]e^{-j\frac{2\pi}{\lambda}\sin\theta\left(nd_x\cos{\phi}+md_y\sin{\phi}\right)}.
\end{split}
\label{eq:beampattern}
\end{align}
Fig. \ref{fig:beampattern} depicts an example of a typical beam pattern created by DAS where we set $w_R[n,m]\equiv1$. The beam pattern can be rewritten as the 2D spatial discrete-time Fourier transform of the aperture function $w_R[\cdot,\cdot]$
\begin{align}
\begin{split}
H_{RX}(\theta,\phi)&=
\sum_{n,m} w_R[n,m]e^{-j s_xn}e^{-j s_ym} \\
&=\mathcal{F}_{2D}\{w_R\}(s_x,s_y),
\end{split}
\label{eq:Fourierbeampattern}
\end{align} 
where we define the spatial frequencies 
\begin{equation}
  s_x\triangleq\frac{2\pi}{\lambda}d_x\sin\theta\cos\phi,\;
  s_y\triangleq\frac{2\pi}{\lambda}d_y\sin\theta\sin\phi.
  \label{eq:sptialfreqs}
\end{equation}
Thus, the design of the beam pattern translates to determining the aperture function.

The final image quality is also affected by the transmit beam pattern, hence, we consider the two-way beam pattern given by the point-wise product

\begin{equation}
    H(\theta,\phi)\triangleq (H_{TX}\cdot H_{RX})(\theta,\phi).
\end{equation}
Here $H_{TX}(\theta,\phi)$ is the transmit beam pattern which by the reciprocal theorem \cite{jensen2002ultrasound} can be written similarly to $H_{RX}(\theta, \phi)$ where we replace $w_R[\cdot,\cdot]$ with the transmit aperture function $w_T[\cdot,\cdot]$.
Thus, the two-way beam pattern can be rewritten as
\begin{equation}
    H(\theta,\phi)=\mathcal{F}_{2D}\{w_T\ast w_R\}(s_x,s_y),
    \label{eq:2Dconvolution}
\end{equation}
where $\ast$ represents a 2D spatial convolution. 

We note that both far-field and narrow-band assumptions, used in the development of \eqref{eq:received signal} and \eqref{eq:delay}, generally do not hold in ultrasound imaging, making the beam pattern expression theoretically invalid. However, it provides a practical tool for assessing the image resolution and contrast, governed by the main lobe and side lobes of the 2D beam pattern\cite{jensen1999linear}.

\subsection{Problem Description} 
\label{subsec:problem}
The main goal of this work is to enable 3D ultrasound imaging with reduced or limited hardware, paving the way for regular use of 3D US in clinics worldwide. We consider the following system aspects: frame rate, image resolution, contrast, and the number of transducer elements on reception. The latter requires cumbersome receive electronics and imposes large computational burden which has an adverse effect on the former system aspects. Throughout the paper, we assume the transducer wavelength $\lambda$ and the element spacing $d_x$ and $d_y$ are given parameters and cannot be changed. Moreover, the array aperture and possible element locations are fixed such that our task of reducing hardware translates to removing some of the elements upon reception.

We introduce a 2D convolutional beamformer which synthetically mimics the convolution operation in (\ref{eq:2Dconvolution}) that occurs naturally due to the physics of the imaging system. This effectively creates a large aperture which in turn leads to improved resolution and contrast. When combined with diverging-wave transmission, our beamforming strategy enables high frame-rate, sufficient for 3D ultrafast imaging. Furthermore, while all elements are utilized for transmission, the proposed beamformer enables the use of sparse arrays, leading to dramatic reduction in the number of receive elements. As the construction of such thinned arrays poses another engineering challenge, we present a scalable 2D sparse array design based on fractals. Here we extend the design recently proposed in \cite{cohen2020sparse} to the 2D setting.

\section{Sparse 3D Convolutional Beamforming}
\label{sec:3Dcoba}
In this section, we present our main contribution: sparse beamforming techniques for ultrafast 3D imaging. We start with a brief description of the concept of the sum co-array followed by the introduction of our 2D convolutional beamformer. As the major computational burden arises from the receive hardware, we perform element reduction by employing 2D sparse arrays upon reception. To complete our proposed framework, we describe a sparse array design based on fractal geometries.
Note that all elements are used for transmission, implying that the transmit beam pattern remains unchanged. Therefore, as we show later, we achieve enhanced image quality by obtaining improved receive beam pattern. 

\subsection{Preliminaries of Array Theory}
We briefly present key concepts of array theory on which convolutional beamforming is based. We start with the following definition.  
\label{subsec:array}
\theoremstyle{definition}
\begin{definition}{Element Set:}
Consider a planar array where $d_x$ and $d_y$ are the minimum spacing in the $xy$ plane of the underlying grid on which sensors are located. The \textit{element set} is defined as an integer set $E$ of tuples where $(n,m)\in E$ if there is a sensor located at $(nd_x,md_y,0)$.
\label{def:positionset}
\end{definition}
\noindent For simplicity, we refer to a planar array with element set $E$ as a planar array $E$. We continue with the definition of the sum co-array.
\theoremstyle{definition}
\begin{definition}{Sum Co-Array:}
Consider a planar array $E$. Define the sum-set of $E$ as 
\begin{equation}
S_E = \big\{(n+u,\,m+v):\quad (n,m),\,(u,v)\in E\big\}.
\end{equation} 
The \textit{sum co-array} of $E$ is defined as the array whose element set is $S_E$, i.e., the planar array $S_E$.
\label{def:sumarray}
\end{definition}

\begin{figure}[h]
 \centering
 \includegraphics[trim={7cm 1cm 7cm 1cm},clip,height = 10cm, width = 0.9\linewidth]{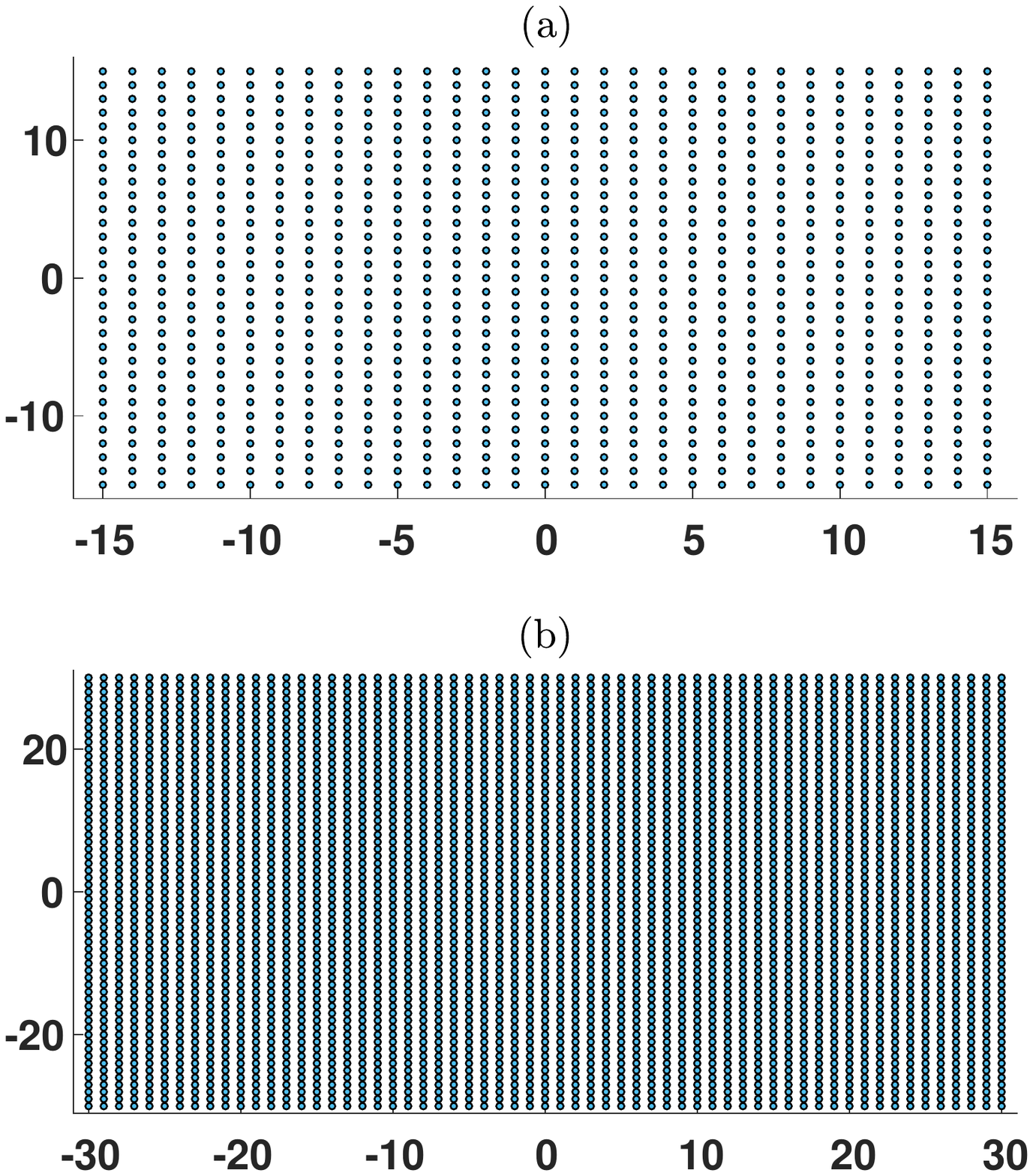}
 \caption{(a) A uniform planar array of $31^2=961$ physical elements and (b) the corresponding sum co-array with $61^2=3721$ virtual elements. Each circle represents an array element.}
  \label{fig:upa}
 \end{figure}

\noindent In Fig.\,\ref{fig:upa}, we show an example of a sum co-array of a uniform planar array (UPA) which is another UPA of twice the size at each axis. 

An additional important part of convolutional beamforming is intrinsic apodization defined below.  

\theoremstyle{definition}
\begin{definition}{Intrinsic Apodization:} 
Consider a planar array $E$ and define a binary indicator matrix $\bf I$ whose entries are ${\bf I}[n,m]=1$ if $(n,m)\in E$ and zero otherwise.
The \textit{intrinsic apodization} of $E$ is an integer matrix defined as 
\begin{equation}
{\bf A}_E\triangleq{\bf I}\ast {\bf I},
\label{eq:intrinsicapodization}
\end{equation}
where $\ast$ denotes 2D convolution operation. Alternatively, the entries $a_E[n,m]$ of ${\bf A}_E$ can be written as
\begin{equation}
\small
    a_E[n,m]= \abs{\Big\{\Big((u,v),(k,l)\Big)\in E^2: u+k=n,v+l=m\Big\}}.
    \label{eq:intrinsicapodizationentries}
\end{equation}
\label{def:intapod}
\end{definition}

The sum co-array and the intrinsic apodization play important roles as they directly affect image quality. As we later demonstrate, the former dictates the size/support of the effective aperture created by COBA-3D, while the latter determines the weights of this aperture.

\subsection{3D Convolutional Beamforming}
\label{subsec:3Dcoba}
Now, we introduce COBA-3D which extends convolutional beamforming \cite{cohen2018coba} to 3D settings with diverging wave transmission. We include an analysis of the consequent sum co-array and receive beam pattern, leading to improved image quality. We note that although the beamforming process is performed digitally, we use in the following the continuous time notation $t$ to simplify the presentation.  

Consider imaging with a UPA $E$ where we transmit a series of $K$ diverging-waves with inclination angles $\{\alpha_k,\beta_k\}_{k=0}^K$ and record the reflected echoes following each transmission. Upon reception, IQ sampling is first performed at a sampling rate at least as high as the Nyquist rate \cite{eldar2015sampling, steinberg1992digital} determined by the transducer center frequency. To achieve a posteriori synthetic transmit focusing, we coherently compound \cite{montaldo2009coherent} the complex received signals by applying to each signal appropriate time delays depending on the element positions and the desired direction $(\theta_0, \phi_0)$. Then, we sum (compound) the results over all transmissions to obtain the compound signal, given by 
\begin{equation}
    y_{n,m}(t)\triangleq \sum_{k=0}^K f_{n,m,\alpha_k,\beta_k}(t+\gamma_{n,m,\alpha_k,\beta_k}),
    \label{eq:compundsignal}
\ifCLASSINFOpdf\end{equation}
where $f_{n,m,\alpha_k,\beta_k}$ denotes the signal received by the element $(n,m)$ following the transmission with inclination angles $\{\alpha_k,\beta_k\}$, and $\gamma_{n,m,\alpha_k,\beta_k}$ is the corresponding time delay (please refer to Chapter I.B of \cite{roux20162d}). Note that the time-point $t$ in \eqref{eq:compundsignal} is proportional to the desired axial depth $z$ via $z=\frac{ct}{2}$.

Next, for any $n\in[-N,N]$ and $m\in[-M,M]$, we define the following signals  
\begin{align}
&r_{n,m}(t)\triangleq \sqrt{|y_{n,m}(t)|}\exp\{j\phase{y_{n,m}(t)}\},
\label{eq:modifiedsignal}
\end{align}
where $|\cdot|$ and $\phase{\cdot}$ are the modulus and phase of the signal respectively.
Denoting by ${\bf r}(t)$ the matrix whose entries are $r_{n,m}(t)$, we define for each time-point the convolution signal
\begin{equation}
{\bf c}(t) = ({\bf r}\ast{\bf r})(t),
\label{eq:convsignal}
\end{equation} 
where $\ast$ denotes 2D spatial convolution.
The matrix ${\bf c}(t)$ is of size $(4N+1)\times(4M+1)$ and can be computed efficiently using a spatial 2D-FFT and its inverse 2D-IFFT as
\begin{equation}
{\bf c}(t) = \text{2D-IFFT}\Big\{\text{2D-FFT}\{{\bf r}\}\odot\text{2D-FFT}\{{\bf r}\}\Big\}(t),
\label{eq:freqconvsignal}
\end{equation} 
where $\odot$ represents the Hadamard product. The signal, beamformed at direction $(\theta_0, \phi_0)$, is then obtained by summing all the entries of ${\bf c}(t)$
\begin{equation}
b(t)\triangleq \sum_{n=-2N}^{2N}\sum_{m=-2M}^{2M} \tilde{w}_R[n,m]{\bf c}_{n,m}(t),
\label{eq:cobasignal}
\end{equation}
where $\{{\bf c}_{n,m}(t)\}$ are the entries of ${\bf c}(t)$ and $\tilde{w}_R[n,m]$ are the weights (apodization) applied to the convolution signal. To attain an effective apodization of $w_R[n,m]$, the actual weights should be set as
\begin{equation}
    \tilde{w}_R[n,m]=\frac{w_R[n,m]}{a_E[n,m]},
    \label{eq:actingweights}
\end{equation}
accounting for the intrinsic apodization $a_E[n,m]$ given by \eqref{eq:intrinsicapodizationentries} that stems from the convolution operation.

The dimension (length) of the vector $b(t)$ and the number of such scan-lines are determined by the discretization along depth (time) and number of chosen directions $(\theta,\phi)$, which represent the underlying grid of the reconstructed image. 
Finally, collecting all beamformed signals $b(t)$ of all directions allows to compose the complete volumetric image defined over a predetermined 3D grid where any region of interest within it is available for visualization.  

We summarize COBA-3D in Algorithm\,\ref{alg:dwcoba}. We note that here we utilize diverging-waves, however, COBA-3D can be performed with focused transmission or any other unfocused insonification such as plane waves. When focused mode is utilized, the first stage of compounding is skipped since focusing is performed upon transmission.  

\begin{algorithm}
\caption{\small COBA-3D}
\label{alg:dwcoba}
{\fontsize{9}{15}\selectfont
\begin{algorithmic} 
\REQUIRE IQ signals $\{f_{n,m,\alpha,\beta}(t)\}$, weights $\{w_R[n,m]\}$. 
\STATE {\bf 1:} Perform coherent compounding
\begin{equation*}
    y_{n,m}(t)= \sum_{k=0}^K f_{n,m,\alpha_k,\beta_k}(t+\gamma_{n,m,\alpha_k,\beta_k})
\end{equation*}
\STATE {\bf 2:} Compute $r_{n,m}(t)= \exp\{j\phase{y_{n,m}(t)}\}\sqrt{|y_{n,m}(t)|}$ 
\STATE {\bf 3:} Perform 2D convolution using 2D FFT
\begin{equation*}
{\bf c}(t) = \text{IFFT}\Big\{\text{FFT}\{{\bf r}\}\odot\text{FFT}\{{\bf r}\}\Big\}(t)
\end{equation*} 
\STATE {\bf 4:} Set weights $\tilde{w}_R[n,m]=\frac{w_R[n,m]}{a_E[n,m]}$
\STATE {\bf 5:} Calculate the beamformed signal
\begin{equation*}
b(t)\triangleq \sum_{n=-2N}^{2N}\sum_{m=-2M}^{2M} \tilde{w}_R[n,m]{\bf c}_{n,m}(t)
\end{equation*}
\ENSURE Beamformed signal $b(t)$.  
\end{algorithmic}}
\end{algorithm}

\begin{figure}[h]
 \centering
 \includegraphics[trim={2cm 3cm 2.5cm 3cm},clip,height = 4cm, width = 0.9\linewidth]{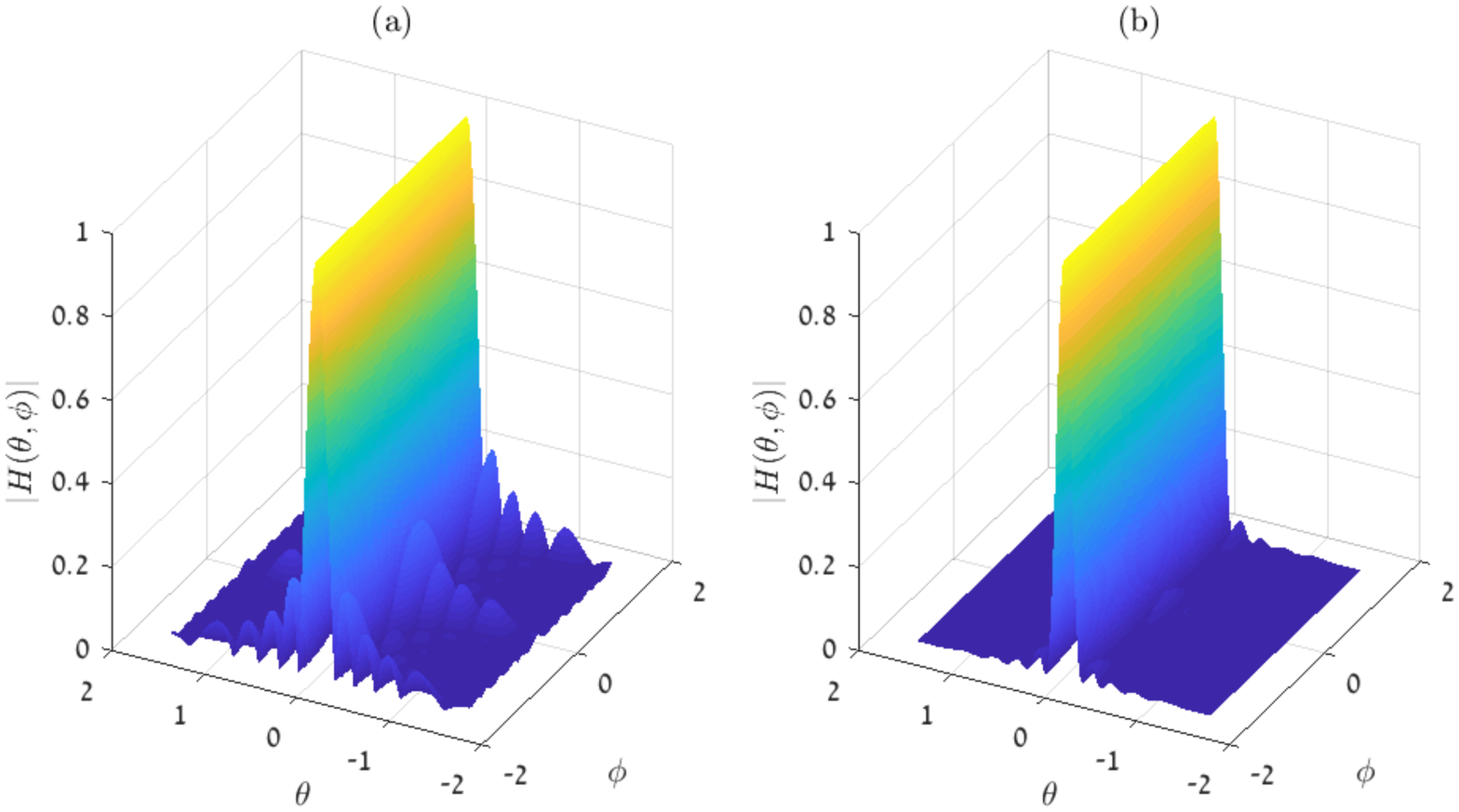}
 \caption{Receive beam patterns: (a) DAS and (b) COBA-3D. Both beamformers use unity receive weights.}
  \label{fig:beampatterncoba}
 \end{figure}

In the Appendix we analyze the receive beam pattern created by COBA-3D, leading to the following expression
\begin{align}
\begin{split}
H_{RX}(\theta,\phi)&=
\sum_{(n,m)\in S_E} w_R[n,m]e^{-j (s_xn+s_ym)}\\
&=\sum_{n=-2N}^{2N}\sum_{m=-2M}^{2M} w_R[n,m]e^{-j (s_xn+s_ym)}
\end{split}
\end{align}
where $w_R[\cdot,\cdot]$ are the effective receive weights. 
As can be seen, the receive beam pattern is directly related to the sum co-array whose aperture is larger than that of the physical array. The latter leads to a receive beam pattern with narrower main lobe and lower side lobes in comparison to DAS as demonstrated in Fig.\,\ref{fig:beampatterncoba}. 
Furthermore, for appropriate choice of the inclination angles, coherent compounding effectively generates a posteriori synthetic focusing in the transmission \cite{montaldo2009coherent}, thus creating a transmit beam pattern $H_{TX}(\theta,\phi)$ that is similar to or even better than that achieved by standard focused transmissions. Thus, COBA-3D leads to an overall beam pattern of  $H_{COBA-3D}(\theta,\phi)= (H_{TX}\cdot H_{RX})(\theta,\phi)$ which is superior to that created by DAS, and thus it should theoretically result in enhanced image quality. This result is verified by practical experiments in Section\,\ref{sec:results}. 

The resultant algorithm COBA-3D resembles COBA presented in \cite{cohen2018coba} but differs from it in three major aspects. First, we add a preprocessing step of compounding which allows COBA-3D to handle unfocused transmissions and increase frame rate. COBA considers focused transmission which limits the frame rate due to the large number of required scan-lines that can reach several tens of thousands in a 3D setup. On the other hand, the frame rate achieved by COBA-3D is dictated by the number of inclination angles $K$ which is typically one order of magnitude lower than the number of scan-lines required in focused imaging, thus enabling real-time imaging. Second, COBA-3D operates on IQ signals which removes the need to perform post band-pass or high-pass filtering. As explained in \cite{cohen2018coba}, when RF data is used the spatial convolution leads to undesired low frequency components that need to be filtered out. The latter side effect is avoided when IQ data is used, thus, removing the post filtering step of COBA. Last and most important, COBA-3D recovers volumetric data and thus reduces operator dependency in the imaging process. Once the 3D data is obtained, any view of any region within it can be displayed without operating the probe, including 
complete views of anatomical body structures that are are intrinsically 3D such as the mitral valve. Moreover, the volume data provides exact location and orientation information, thus, a variety of parameters can
be estimated from a 3D image in a more accurate and reproducible fashion compared to 2D imaging \cite{prager2010three, abo2004usefulness, kwan2014three}.
 
To conclude this part, COBA-3D allows to handle diverging wave transmissions, thus, obtaining ultrafast frame rate. In addition, it generated an improved receive beam pattern which potentially should lead to higher image quality than DAS. A challenge that remains is the heavy receive hardware and the corresponding sizable data that needs to be processed upon reception due to the large number of receive elements.    


\subsection{Sparse Beamforming}
\label{subsec:3Dscoba}

So far, we introduced COBA-3D which is designed to achieve ultrafast frame rate and improved image resolution and contrast compared to DAS. However, still a large number of receive elements is required, leading to high cost, power and computational burden. Moreover, the latter increases considerably when transmitting diverging-waves, since the entire region of interest is reconstructed following each transmission (and not a single scan-line). 

To obtain element reduction without degrading performance in terms of image resolution and contrast, we utilize 2D sparse arrays defined next. This is a generalization of the sparse arrays introduced in \cite{cohen2018coba} to the 2D setting.

\theoremstyle{definition}
\begin{definition}{2D Sparse Array:}
Let $E$ and $T$ be two planar arrays, and denote by $S_T$ the sum co-array of $T$. We say $T$ is a \textit{sparse (or thinned) array} with respect to $E$ if
\begin{equation}
T\subset E \subseteq S_T,
\label{eq:sparsedefinition}
\end{equation}  
where in the above we consider the elements sets of the arrays.  
\label{def:sa}
\end{definition}  

 Suppose we perform imaging by using a UPA $E$ for transmitting tilted diverging-waves where upon reception we employ a thinned (sparse) array $T$ for acquiring the backscattered signals. We can no longer compute $b(t)$ as in \eqref{eq:convsignal} or \eqref{eq:freqconvsignal} since we removed some of the elements (or signals). Therefore, we compute the convolution signal $\textbf{c}(t)$ using pair-wise multiplications of the signals as follows. Denoting by $S_T$ the sum co-array of $T$, for any $(n,m)\in S_T$ we define  
\begin{equation}
    c_{n,m}(t) = \sum_{(u,v)\in T}\; \sum_{\substack{(k,l)\in T:\\ u+k=n \\ v+l=m}} (r_{u,v}\cdot r_{k,l})(t).
    \label{eq:cnm}
\end{equation}
Alternatively, we can obtain (\ref{eq:cnm}) by filling in the missing signals with zeros and then performing a 2D convolution
\begin{equation}
    {\bf c}(t) = (\tilde{\bf r}\ast\tilde{\bf r})(t),
    \label{eq:cnmconv}
\end{equation}
where
\begin{equation}
    \tilde{r}_{n,m}(t)=
    \begin{cases}
    r_{n,m}(t),\;&(n,m)\in T, \\
    0,\;&\text{otherwise.}
    \end{cases}
\end{equation}
Finally, the beamformed signal is given by 
\begin{equation}
    b(t)=\sum_{(n,m)\in S_T}  \tilde{w}_R[n,m]c_{n,m}(t),
    \label{eq:scobasignal}
\end{equation}
where we incorporated the receive aperture function $\tilde{w}_R[n,m]$, determined by considering the intrinsic apodization of the sparse array $T$ as in \eqref{eq:actingweights}. Again, this process yields a single scan line $b(t)$ of a specific direction and should be repeated for all desired directions to obtain the complete 3D image. The resultant technique, referred as sparse COBA-3D (SCOBA-3D), is outlined in Algorithm\,\ref{alg:dwscoba}. 

\begin{algorithm}
\caption{\small SCOBA-3D}
\label{alg:dwscoba}
{\fontsize{9}{15}\selectfont
\begin{algorithmic} 
\REQUIRE IQ signals $\{f_{n,m,\alpha,\beta}(t)\}$, weights $\{w_R[n,m]\}$. 
\STATE {\bf 1:} Perform coherent compounding
\begin{equation*}
    y_{n,m}(t)= \sum_{k=0}^K f_{n,m,\alpha_k,\beta_k}(t+\gamma_{n,m,\alpha_k,\beta_k})
\end{equation*}
\STATE {\bf 2:} Compute $r_{n,m}(t)= \exp\{j\phase{y_{n,m}(t)}\}\sqrt{|y_{n,m}(t)|}$
\STATE {\bf 3:} Calculate $c_{n,m}(t)$ using (\ref{eq:cnm}) or (\ref{eq:cnmconv}) for all $(n,m)\in S_T$
\STATE {\bf 4:} Set weights $\tilde{w}_R[n,m]=\frac{w_R[n,m]}{a_T[n,m]}$ for all $(n,m)\in S_T$
\STATE {\bf 5:} Compute the beamformed signal using (\ref{eq:scobasignal})
\begin{equation}
    b(t)=\sum_{(n,m)\in S_T}  \tilde{w}_R[n,m]c_{n,m}(t)
\end{equation}
\ENSURE Beamformed signal $b(t)$.  
\end{algorithmic}}
\end{algorithm}

Following the same steps as in the Appendix, we can obtain an expression of the receive beam pattern created by SCOBA-3D
\begin{equation}
H_{RX}(\theta,\phi)=
\sum_{(n,m)\in S_T} w_R[n,m]e^{-j (s_xn+s_ym)}
\label{eq:scobabeampattern}
\end{equation}
where $w_R[n,m]$ are the effective apodization and $a_T[n,m]$ are the intrinsic apodization of $T$ defined in \eqref{eq:intrinsicapodizationentries}. By Definition\,\ref{def:sa}, we have that $E\subset S_T$, hence, we can write
\begin{align}
\begin{split}
H_{RX}(\theta,\phi)&=
\sum_{(n,m)\in E} w_R[n,m]e^{-j (s_xn+s_ym)}\\&+\sum_{(n,m)\in S_T/E} w_R[n,m]e^{-j (s_xn+s_ym)},
\end{split}
\end{align}
where $S_T/E\triangleq\left\{(n,m)\in S_T:\,(n,m)\notin E\right\}$. Therefore, condition (\ref{eq:sparsedefinition}) ensures that the resultant sum co-array exhibits at least as large aperture as that of the fully-populated array $E$. In the special case where we set $w_R[n,m]=0$ for all $(n,m)\in S_T/E$, expression (\ref{eq:scobabeampattern}) reduces to 
\begin{equation}
    H_{RX}(\theta,\phi)=
\sum_{(n,m)\in E} \hat{w}_R[n,m]e^{-j (s_xn+s_ym)},
\end{equation}
which is the receive beam pattern achieved by DAS operating on the full array $E$. Hence, SCOBA-3D provides more degrees of freedom in choosing the apodization weights than DAS.
Finally, since we remove elements only upon reception, the transmit beam pattern remains unchanged, as before.

\begin{figure*}[htb]
 \centering
 \includegraphics[trim={2cm 3cm 2cm 3cm},clip,height = 7cm, width = 0.9\linewidth]{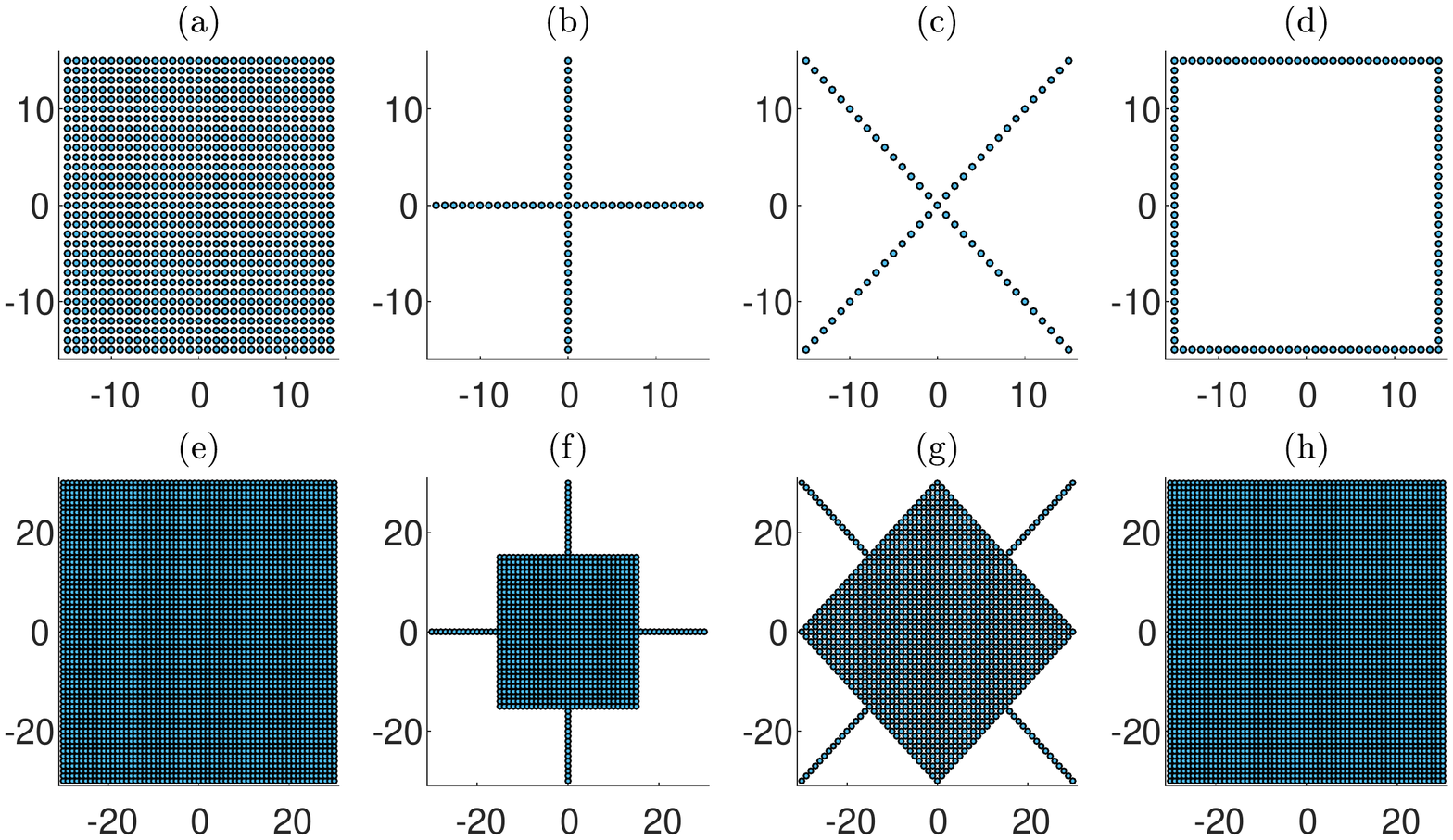}
 \caption{Sparse Arrays: (a) the original UPA - 961 elements, (b) '$+$'-shape array - 61 elements, (c) 'X'-shape array - 61 elements , (d) '$\Box$'-shape array - 120 elements. (e)-(h) are the corresponding sum co-arrays with  3721, 1021, 1021 and 3721 elements respectively. Each circle represents an array element. All sum co-arrays contains within them a UPA which in turn includes the original UPA.}
  \label{fig:arraysbeampatten}
 \end{figure*}

\begin{figure*}[htb]
 \centering
 \includegraphics[trim={0cm 5cm 2cm 4cm},clip,height = 6cm, width = 0.9\linewidth]{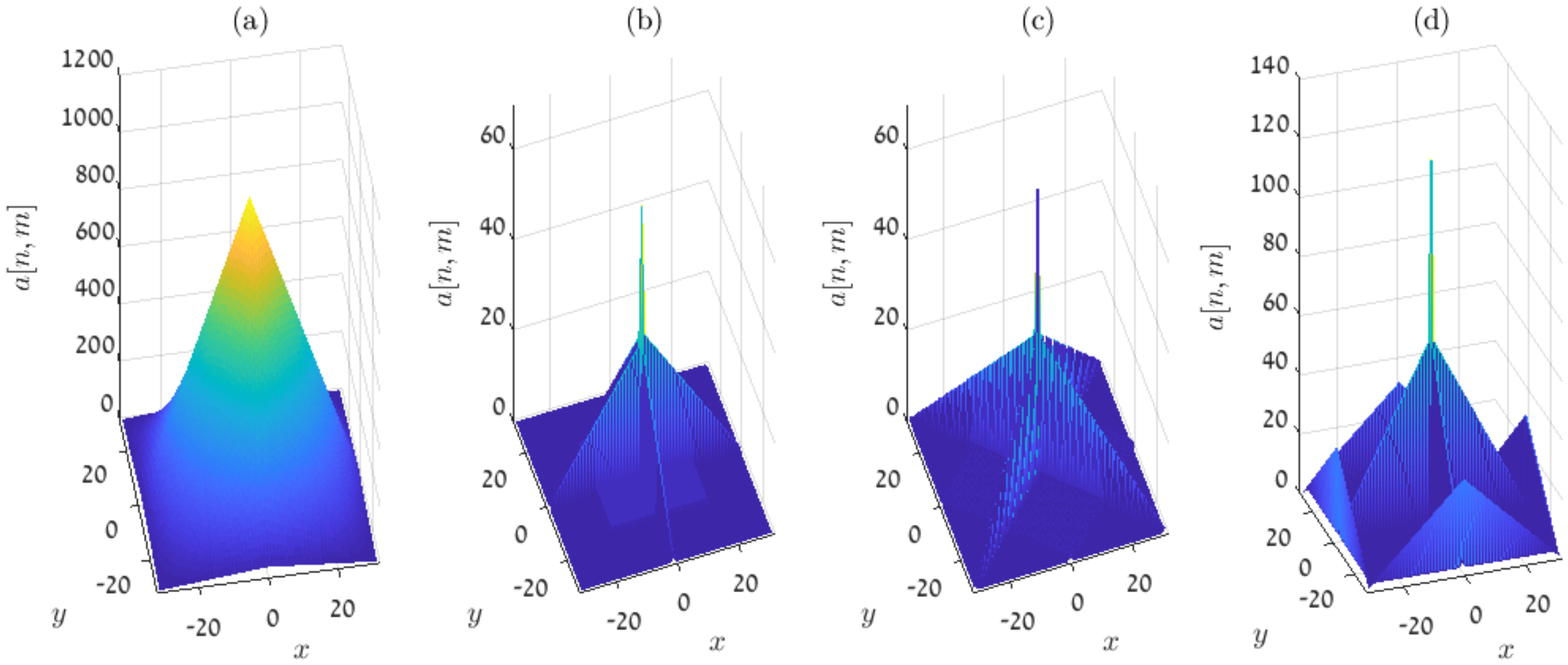}
 \caption{Intrinsic apodization of various arrays: (a) UPA, (b) '$+$'-shape array, (c) 'X'-shape array, (d) '$\Box$'-shape array. Any choice of sparse array leads to different intrinsic apodization that can improve or degrade image quality. Hence, the intrinsic apodization should be considered when setting the receive weights.}
  \label{fig:intrinsic}
 \end{figure*}

Fig.\,\ref{fig:arraysbeampatten} presents examples of various sparse arrays and their corresponding sum co-arrays. All sum co-arrays in this case includes a UPA within them. A special case is the 'X'-shape array whose sum co-array contains a tilted UPA. As seen, the use of sparse geometries offers a dramatic reduction in the number of elements, leading to a number of elements typically used for 2D imaging. In addition, different sparse arrays lead to different intrinsic apodization weights as shown in Fig.\,\ref{fig:intrinsic}. Some intrinsic apodization functions can improve image quality, e.g. Fig.\,\ref{fig:intrinsic}(a) which reduces side lobes, while others exhibit discontinuities due to zero weights that might lead to an adverse effect on the beam pattern. Therefore, any choice of receive weights has to consider the intrinsic apodization caused by the spatial convolution.

\subsection{Fractal Array Design}
\label{subsec:fractals}
Sparse arrays play a major role in convolutional beamforming. Besides the number of physical elements, there are additional important design criteria for sparse arrays that affect the performance. Examples are:

\begin{criterion}[Closed-form]
To allow scalability, elements locations should be given in closed-form.
\label{criterion:closedform}
\end{criterion}

\begin{criterion}[Symmetric Array]
Consider a planar array $T$ and denote by $\hat{T}$ 
a version of the array rotated by $180^{\circ}$:
\begin{equation*}
\hat{T}\triangleq\{(-n,-m)\,|\;(n,m)\in T\}.
\end{equation*}
A planar array $T$ is symmetric if $T=\hat{T}$.
\label{criterion:symmetry}
\end{criterion}

\begin{criterion}[Full Sum Co-Array]
Consider a planar array $T$ whose sum co-array is $S_T$. The sum coarray $S_T$ is said to be \textit{full} (i.e. contiguous) if it is a UPA. This ensures that the aperture of the sum co-array does not exhibit discontinuities which may have adverse effect on the beam pattern. 
\label{criterion:holefree}
\end{criterion}

\begin{criterion}[Large Sum Co-Array]
To obtain considerable element reduction, the sum co-array $S_T$ of a sparse array $T$ should satisfy $\abs{S_T}=\mathcal{O}(\abs{T}^2)$ \cite{hoctor1990unifying}. This implies that the aperture size of the sum co-array is large, leading to improved resolution and contrast. 
\label{criterion:large}
\end{criterion}

\noindent Depending on the specific application, one may consider additional array properties of interest such as mutual coupling (element cross-talk) \cite{cohen2019sparse}, but for simplicity we focus here on the criteria mentioned above, as we did in \cite{cohen2019sparse}.

When considering these criteria, the design of sparse arrays becomes intractable in large scale, i.e., when the number of elements is large as in 3D imaging. To address this issue, we adopt recent work \cite{cohen2020sparse,cohen2019sparse} and extend it to introduce a 2D fractal array design based on the sum co-array.      

\begin{figure*}[htb]
 \centering
 \includegraphics[trim={3cm 2cm 3cm 2cm},clip,height = 6cm, width = 0.6\linewidth]{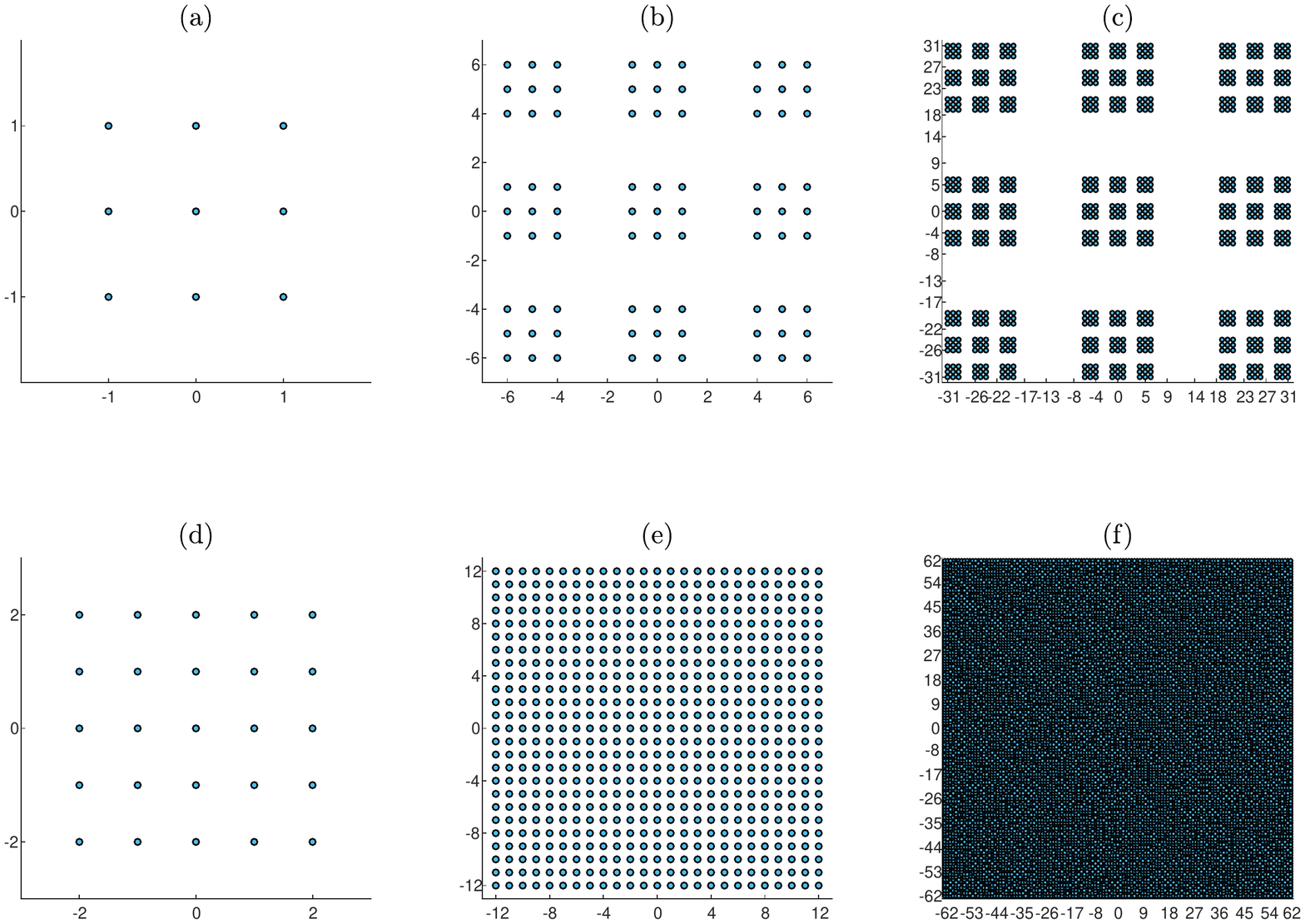}
 \caption{Fractal Arrays: (a) The generator array with 9 elements and its fractal extensions (b) $F_2$ and (c) $F_3$ with 81 and 729 elements respectively. The corresponding sum co-arrays are given in (d), (e) and (f) with 25, 625 and 15625 elements respectively.}
  \label{fig:fractals}
 \end{figure*}

Consider a planar array $T$ whose sum co-array $S_T$ is assumed to be full. We propose the following recursive array definition for any natural number $r$
\begin{align}
\begin{split}
&{F}_0 \triangleq \{0\}, \\
&{F}_{r+1} \triangleq \bigcup_{(n,m)\in{T}} \left({F}_r+(n\cdot 
C_x^r,\,m\cdot C_y^r)\right),
\end{split}
\label{eq:fractal}
\end{align}
where $C_x$ and $C_y$ are the number of elements in each row and column of $S_T$, respectively. Note that $T$, referred to as the generator \cite{cohen2020sparse}, satisfies $T=F_1$. Each fractal array $F_{r+1}$ is composed of $\abs{T}$ replicas of $F_r$ arranged in space according to $T$, leading to a total number of elements of $\abs{T}^r$ in $F_r$.
Fig.\,\ref{fig:fractals} exemplifies the proposed fractal design where we choose the generator to be a UPA. 

In \cite{cohen2020sparse}, a similar fractal array definition was proposed based on the difference co-array rather than the sum co-array as in our scheme. However, for symmetric arrays the sum and difference co-arrays are identical. Therefore, for a symmetric generator, the theoretical proofs derived in \cite{cohen2020sparse} apply to our case, implying that the fractal arrays inherit the properties of their generator. In particular, whenever the generator satisfies Criteria\,\ref{criterion:symmetry}-\ref{criterion:large} so do its fractal expansions.         
Thus, the fractal scheme (\ref{eq:fractal}) provides a simple systematic way for designing sparse arrays by constructing a generator array with desirable properties (using e.g. exhaustive search) and then enlarging it recursively while preserving its properties, as shown in Fig.\,\ref{fig:fractals}.

\section{Evaluation Results}
\label{sec:results}
Here we study the performance of 3D convolutional beamforming operating on thinned receive arrays and compare it to standard DAS applied on the full array. We present experiments performed with the parallelized Verasonics systems at the University of Lyon \cite{petrusca2018fast} where we consider either focused transmission scheme or diverging-wave compounding. First, we present images of phantom scans and examine them to assess image resolution and contrast qualitatively and quantitatively. Then, we show results obtained from \textit{ex-vivo} data to strengthen our validation. In the following, we refer in this section for brevity to COBA-3D and SOCBA-3D as COBA and SCOBA respectively. We begin with a full description of our experimental setup. 


\subsection{System description}
\label{subsec:system}
The acquisition system consists of four Vantage 256 systems (Verasonics, USA) which are synchronized together with an external box (Verasonics, USA). Such a configuration allows controlling 1024 individual channels in both transmission and reception. A 2D probe is connected to the systems, each of them driving 256 elements. The probe is composed of 32 $\times$ 35 elements with a 300~$\mu$m pitch in both $x$ and $y$ direction (Vermon, France). We note that in the $y$ direction, elements line \#9, \#17, and \#25 are not connected~\cite{petrusca2018fast}. For both focused and diverging-wave transmissions, a 2-cycle 3-MHz sinusoidal wave is transmitted into the medium. The reception is conducted with a 12~MHz sampling frequency. 

The difference between the two transmission schemes concerns the position of the focal spot, which is a positive $z$ value in the focalized transmissions and a negative $z$ value for diverging-waves. For the focalized emissions, the focal spot is initially set at 40~mm depth. Then, steering is applied in both elevation ($xz$ plane) and azimuth ($yz$ plane). In both directions, the angle value are in the range [-30$^\circ$; 30$^\circ$], discretized in 49 and 51 angles for elevation and azimuth direction, respectively. This leads to 2499 transmission/reception events. For diverging-waves, the virtual focal spot is located at -4.8~mm, which is half the probe aperture. Then, steering is applied in both planes in the range [-10$^\circ$; 10$^\circ$] with a discretization of 9 angles in both direction, leading to a total of 81 transmission/reception events. 

\begin{figure*}[htb]
\centering
\begin{subfigure}{0.9\textwidth}
  \centering
  \includegraphics[trim={4cm 7cm 2cm 6cm},clip, height = 5cm, width = 1\linewidth]{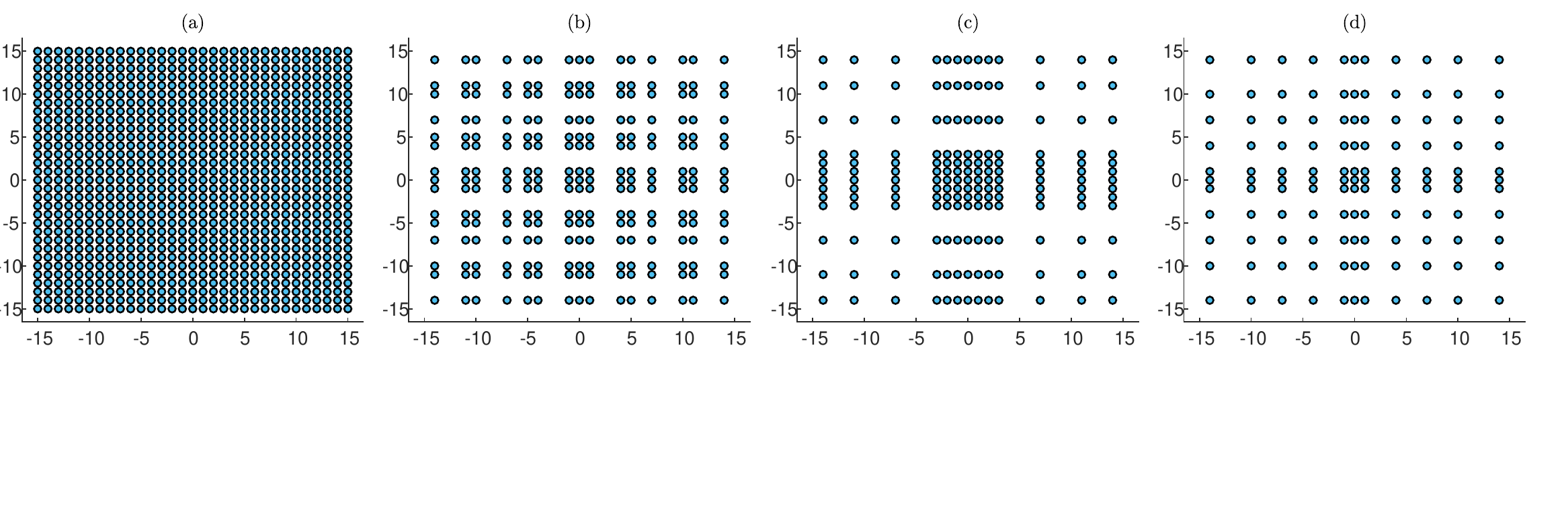}
\end{subfigure}%
\vspace{-1.5cm}
\begin{subfigure}{0.9\textwidth}
  \centering
  \includegraphics[trim={4cm 7cm 2cm 6cm},clip, height = 5cm, width = 1\linewidth]{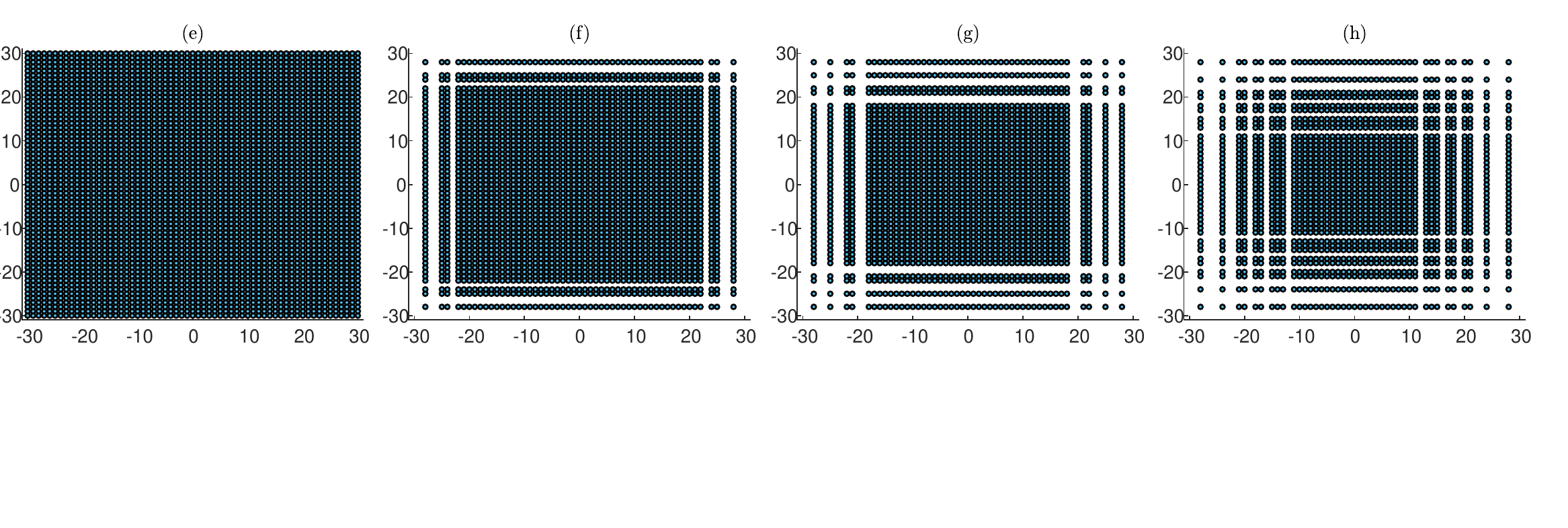}
\end{subfigure}%
\vspace{-1.5cm}
\begin{subfigure}{0.9\textwidth}
  \centering
  \includegraphics[trim={3cm 9cm 2cm 6cm},clip, height = 3cm, width = 1\linewidth]{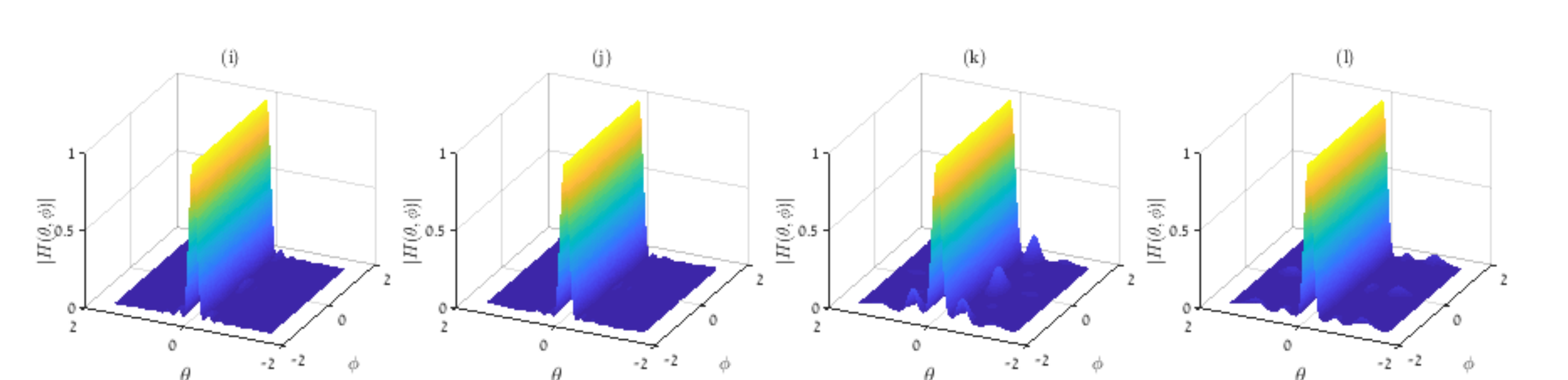}
\end{subfigure}%
\caption{Array Configurations: (a) UPA with 961 elements, (b)-(d) sparse arrays with 225, 169 and 121 elements respectively. The corresponding sum co-arrays are shown in (e)-(h) and the respective beam patterns, exhibited by SCOBA, are given in (i)-(l).}
\label{fig:arrays}
\end{figure*}

Throughout the experiments we use the full aperture for transmission, whereas upon reception we utilize sparse arrays by removing (ignoring) part of the elements out of the full $31\times31$ aperture. The chosen arrays, shown in Fig.\,\ref{fig:arrays}, are non-fractal arrays based on nested arrays \cite{pal2010nested, cohen2018coba} and were selected since they are simple to construct and their sum co-arrays contain a UPA. Moreover, the different thinned configurations allow to  easily display various levels of element reduction and the effect on the size of the UPA contained in the corresponding sum co-arrays. The sum co-arrays lead to different beam patterns as given in Fig.\,\ref{fig:arrays}, leading to different image qualities as we later demonstrate in this section. For clarity, the sparse arrays (b), (c) and (d) of Fig.\,\ref{fig:arrays} are denoted as Array I, II and III respectively. 
We apply DAS and COBA on the UPA of Fig.\,\ref{fig:arrays} and SCOBA on Arrays I, II and III where we refer to the resultant methods as SCOBA I, II and III accordingly. In addition, no apodization is applied in any of the methods examined for fair comparison.  

The proposed fractal design aims at constructing large sparse array. Here, due to our available system setup, we are confined to a $31\times31$ aperture that is considered to be small for our purposes, thus, limiting us in showing the full extent of our recursive array scheme. However, for completeness of our work, we provide additional results obtained using the sparse fractal array in Fig.\,\ref{fig:fractals}(b) which consists of 81 out of 169 ($13\times13$) elements comprising the full counterpart array. Note that this fractal array satisfies Criteria \ref{criterion:closedform}-\ref{criterion:holefree}, but not Criterion\,\ref{criterion:large} since its generator fails to meet it, as seen from Fig.\,\ref{fig:fractals}(a) and Fig.\,\ref{fig:fractals}(d). As a proper comparison, we present results obtained with DAS operating on the full $13\times13$ array with either focused or diverging wave transmission.

\subsection{Phantom Experiments}
\label{subsec:phantom}

\begin{figure*}
 \centering
  \includegraphics[trim={3cm 5cm 2cm 6cm},clip,height = 7cm, width = 0.9\linewidth]{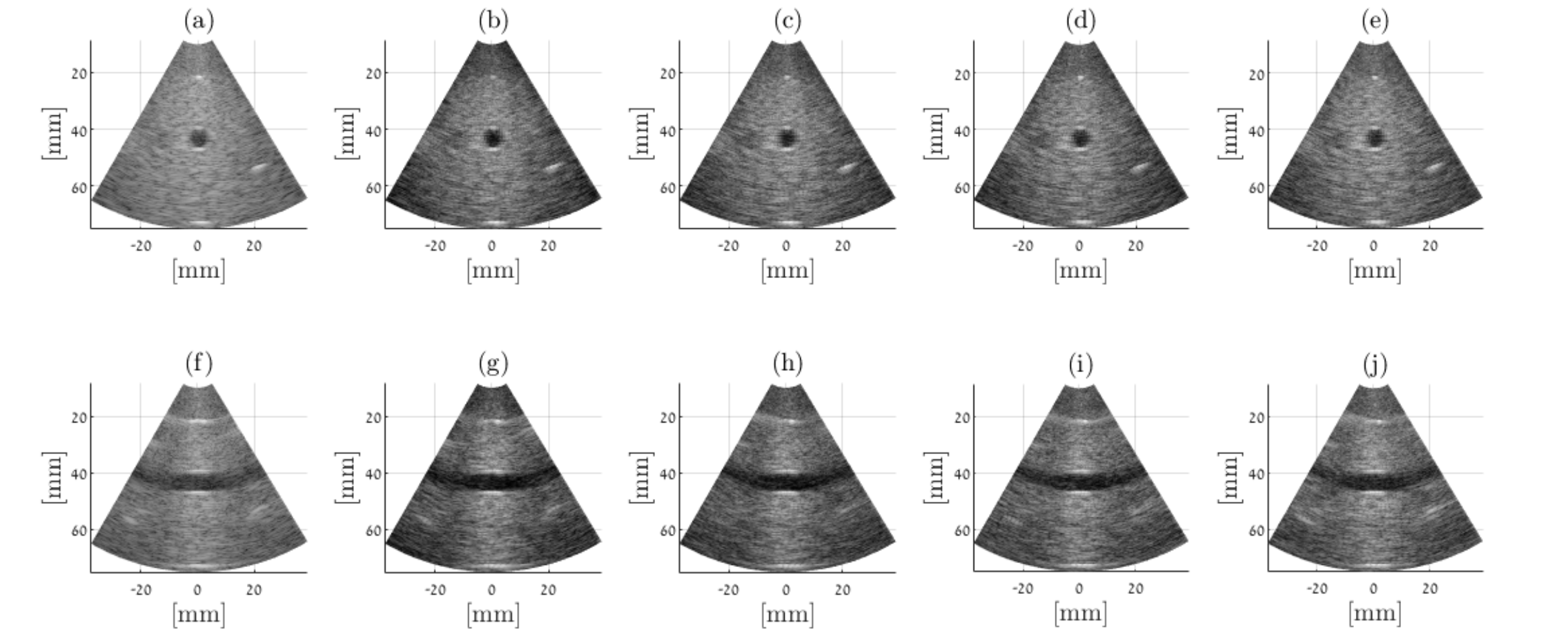}
 \caption{Orthogonal views of (top) $xz$ and (bottom) $yz$ planes  of a Gammex phantom with anechoic cyst obtained using focused transmission with DAS (961), COBA (961), SCOBA I (225), SCOBA II (169), and SCOBA III (121) in their respective order from left to right. Numbers in brackets refer to the number of receive elements. Images are displayed with a dynamic range of 90dB.}
  \label{fig:focus_phantom_contrast}
 \end{figure*}
\begin{figure*}
\centering
\begin{subfigure}{.35\textwidth}
  \centering
  \includegraphics[trim={6cm 3cm 6cm 4cm},clip, height = 4cm, width = 0.8\linewidth]{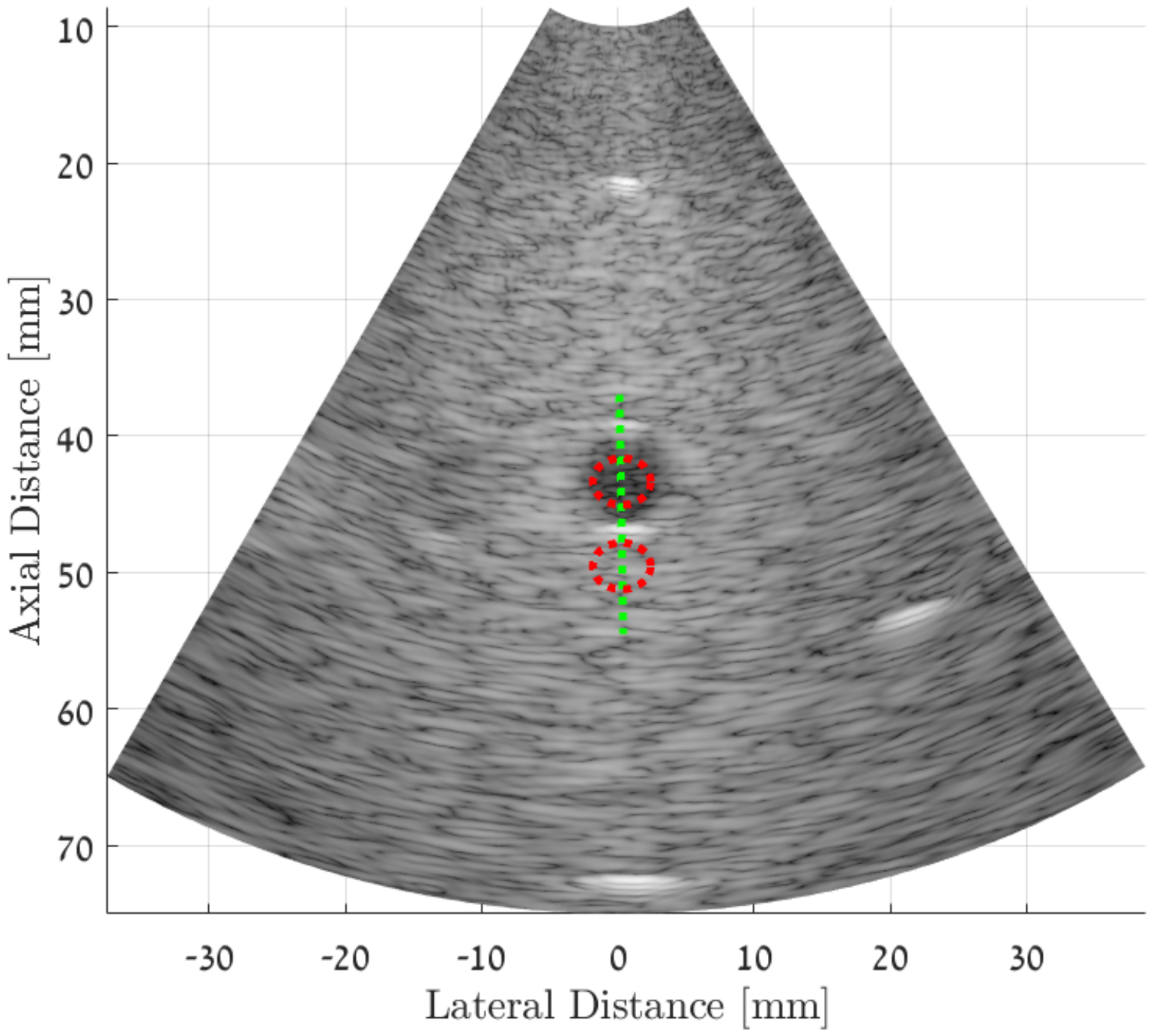}
  \caption{}
\end{subfigure}%
\begin{subfigure}{.55\textwidth}
  \centering
  \includegraphics[trim={2cm 3.5cm 1cm 4cm},clip, height = 4cm, width = 0.9\linewidth]{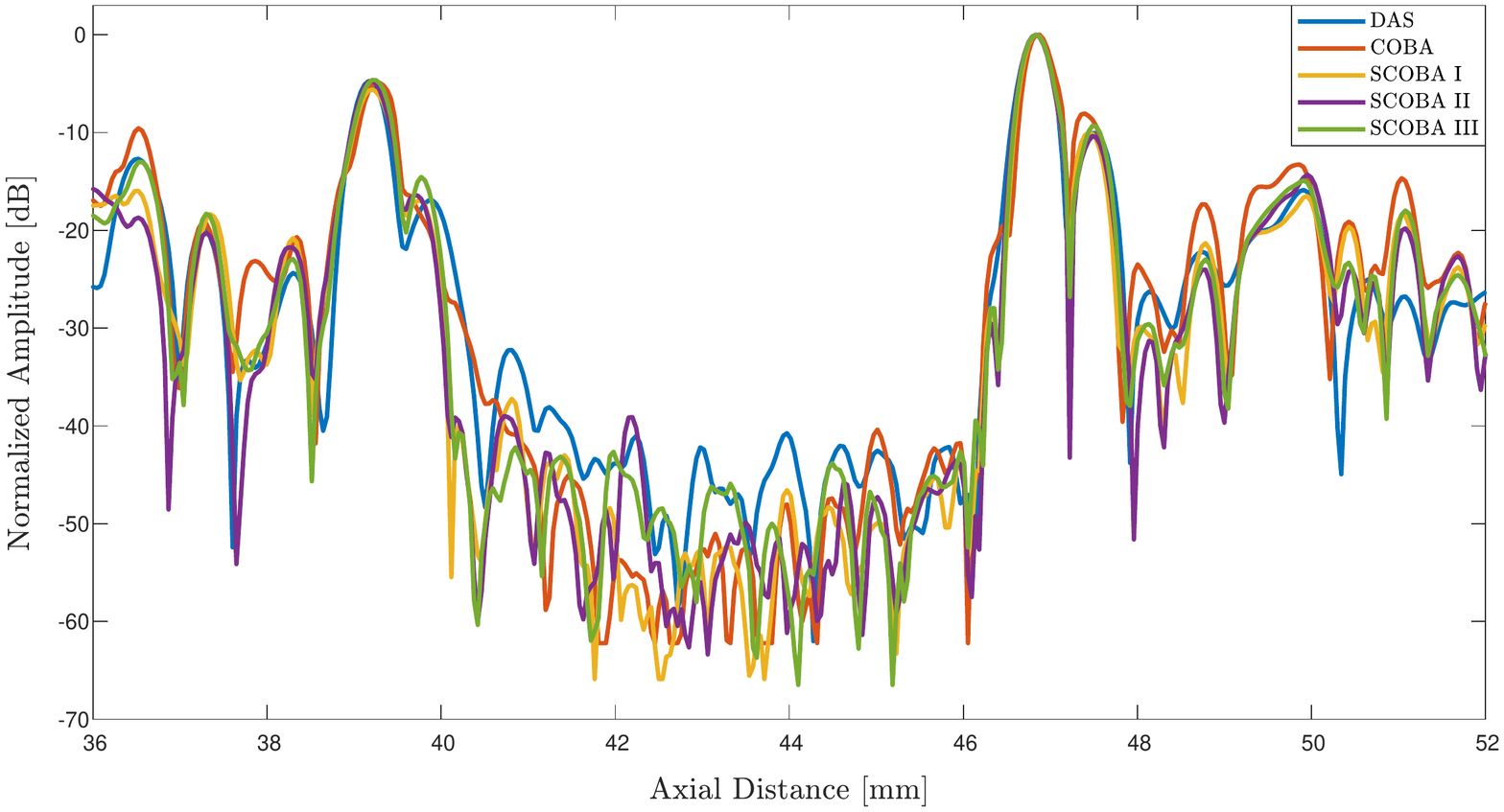}
  \caption{}
\end{subfigure}
\caption{(a) The red circles mark the areas used for computing the CR values of Table\,\ref{tab:contrast}, while the dashed green line indicates the location of the cross-section shown in (b) to evaluate contrast.}
\label{fig:focus_contrast}
\end{figure*}

First, both transmission schemes are used on a commercial grayscale phantom (Gammex Sono410 model). We start with focused transmission. Fig.\,\ref{fig:focus_phantom_contrast} provides various images obtained by the different beamforming techniques. The images include a anechoic cyst which is used to visually assess the contrast of the recovered images. Examining the cyst, we can see that COBA generates sharper images than DAS. SCOBA achieves similar or improved contrast in comparison to DAS while offering 4-8 fold element reduction upon reception. A closer look at the performance of the beamformer is shown in Fig.\,\ref{fig:focus_contrast} where we display cross-sections of the cyst to visually assess contrast. As a quantitative measure, we compute the contrast ratio (CR) \cite{lediju2011short} 
\begin{equation}
\text{CR}=20\log_{10}\left(\frac{\mu_{\text{cyst}}}{\mu_{bck}}\right)
\end{equation}
where $\mu_{\text{cyst}}$ and $\mu_{\text{bck}}$ are the mean image intensities, prior to log-compression, computed over two regions inside the cyst and in the surrounding background, respectively. The chosen regions are marked by dashed red circles in Fig.\,\ref{fig:focus_contrast}. The results, given in Table\,\ref{tab:contrast}, show that COBA exhibits enhanced image contrast as well as SCOBA variants comapred to DAS

\begin{table}[h]
  \begin{center}
    \caption{ \small CR [dB]. \newline The numbers in brackets denote the number of receive elements.} 
    \label{tab:contrast}
    \begin{tabular}{|c|c|c|c|c|c|}
      \hline 
     & \shortstack{\textbf{\small DAS} \\ (961)} & \shortstack{\textbf{COBA} \\ (961)} & \shortstack{\textbf{SCOBA I} \\ (225)} & \shortstack{\textbf{SCOBA II} \\ (169)} & \shortstack{\textbf{SCOBA III} \\ (121)} \\
      \hline 
      \textbf{Focused} & -23.97 & -30.60 & -30.57 & \textbf{-30.71} & -27.86 \\
      \hline 
     \textbf{ DW}   & -7.38  & \textbf{-10.20} & -9.70 & -8.69 & -6.52\\
      \hline 
    \end{tabular}
  \end{center}
\end{table}

\begin{figure*}
 \centering
 \includegraphics[trim={3cm 5cm 2cm 6cm},clip,height = 7cm, width = 0.9\linewidth]{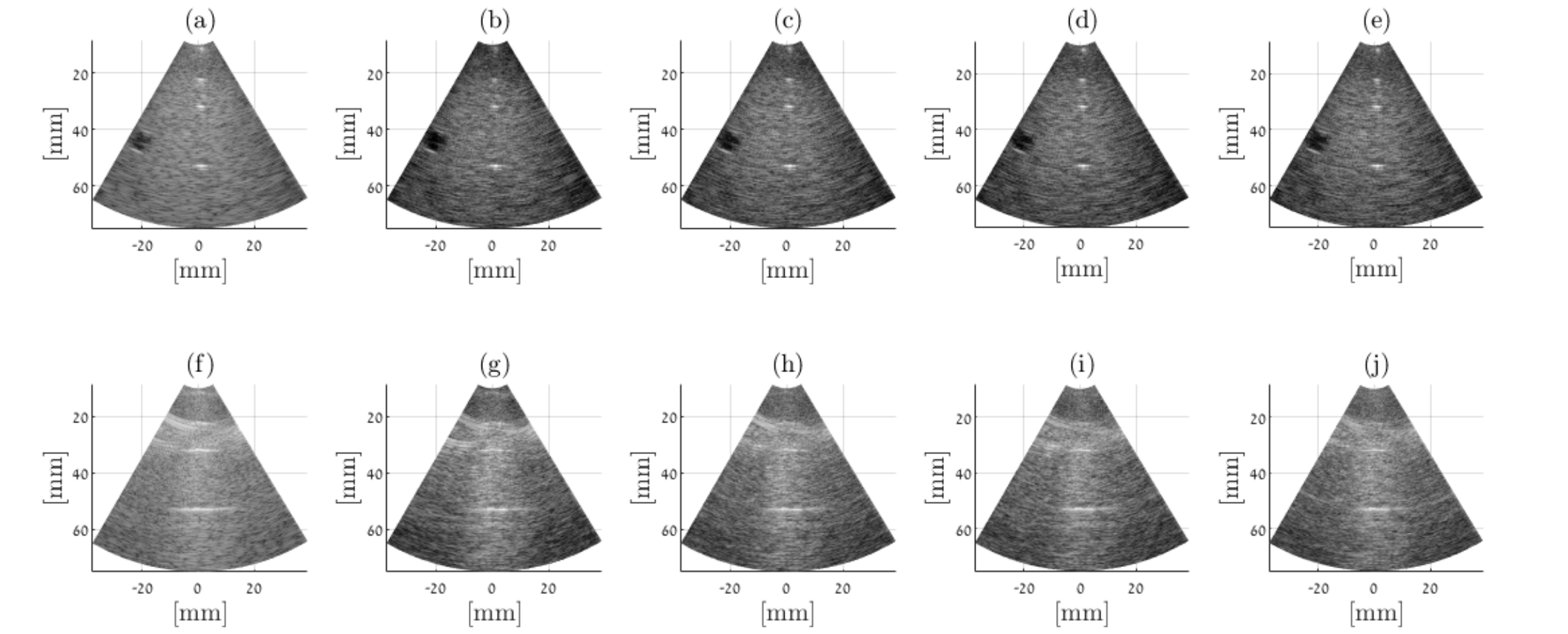}
 \caption{Orthogonal views of (top) $xz$ and (bottom) $yz$ planes of a Gammex phantom with punctual targets obtained using focused transmission with (a) DAS (961), (b) COBA (961), (c) SCOBA I (225), (169) SCOBA II and SCOBA III (121). Numbers in brackets refer to the number of receive elements. Images are displayed with a dynamic range of 90dB.}
  \label{fig:focus_phantom_resolution}
 \end{figure*}

\begin{figure*}
\centering
\begin{subfigure}{.35\textwidth}
  \centering
  \includegraphics[trim={0cm 4.4cm 0cm 5cm},clip, height = 4cm, width = 0.7\linewidth]{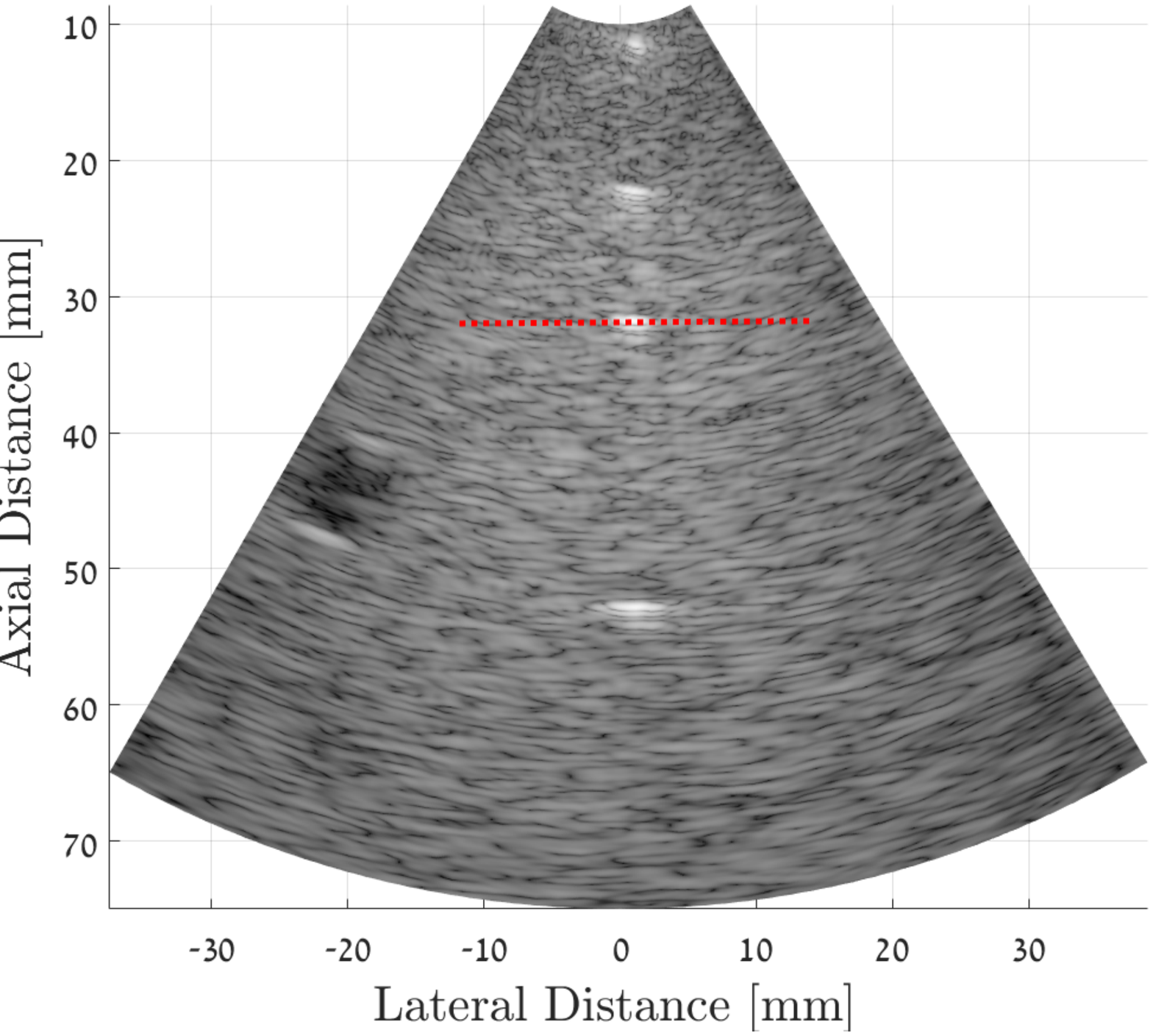}
  \caption{}
\end{subfigure}%
\begin{subfigure}{.55\textwidth}
  \centering
  \includegraphics[trim={3cm 3.5cm 2cm 4.5cm},clip, height = 4cm, width = 0.9\linewidth]{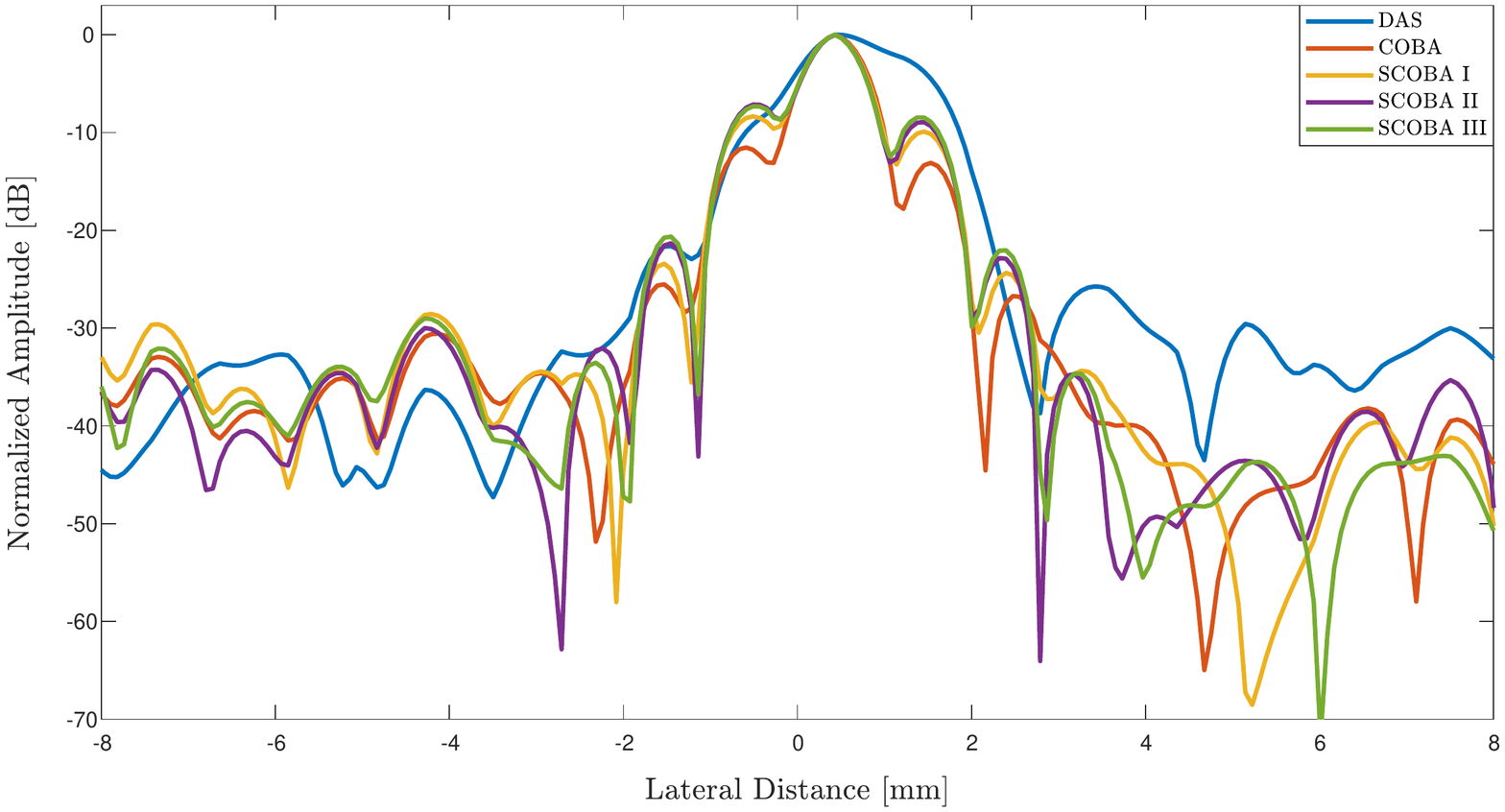}
  \caption{}
\end{subfigure}
\caption{(a) The dashed red line marks the location of the cross-section shown in (b) to assess resolution.}
\label{fig:focus_resolution}
\end{figure*}

Now, we study the resolution of the obtained images. To that end, we use phantom images which comprise point targets. Examining the results in in Fig\,\ref{fig:focus_phantom_resolution}, specifically the points scatterers, we can see that COBA and SCOBA variants provide improved resolution compared to DAS. This is result strengthens when cross-section of a chosen point scatterer is displayed in Fig.\,\ref{fig:focus_resolution}, showing the variants of SCOBA yield better resolution than DAS and COBA outperforms them all. In addition, we compute the full-width at half-maximum (FWHM) to quantitatively measure the resolution obtained by each method. The results are given in Table\,\ref{tab:resolution}, emphasizing the enhanced resolution offered by the proposed beamforming algorithms.    

 \begin{table}[h]
  \begin{center}
    \caption{\small FWHM [mm]. \newline The numbers in brackets denote the number of receive elements.} 
    \label{tab:resolution}
    \begin{tabular}{|c|c|c|c|c|c|}
      \hline 
      & \shortstack{\textbf{\small DAS} \\ (961)} & \shortstack{\textbf{COBA} \\ (961)} & \shortstack{\textbf{SCOBA I} \\ (225)} & \shortstack{\textbf{SCOBA II} \\ (169)} & \shortstack{\textbf{SCOBA III} \\ (121)} \\
      \hline 
      \textbf{Focused} & 1.89  & \textbf{0.94} &  \textbf{0.94} &  \textbf{0.94} &  \textbf{0.94} \\
      \hline 
     \textbf{ DW}   & 2.65  & \textbf{1.36} & 1.68 & 1.76 & 1.84\\
      \hline 
    \end{tabular}
  \end{center}
\end{table}

\begin{figure*}
 \centering
 \includegraphics[trim={3cm 4cm 2cm 6cm},clip,height = 7cm, width = 0.9\linewidth]{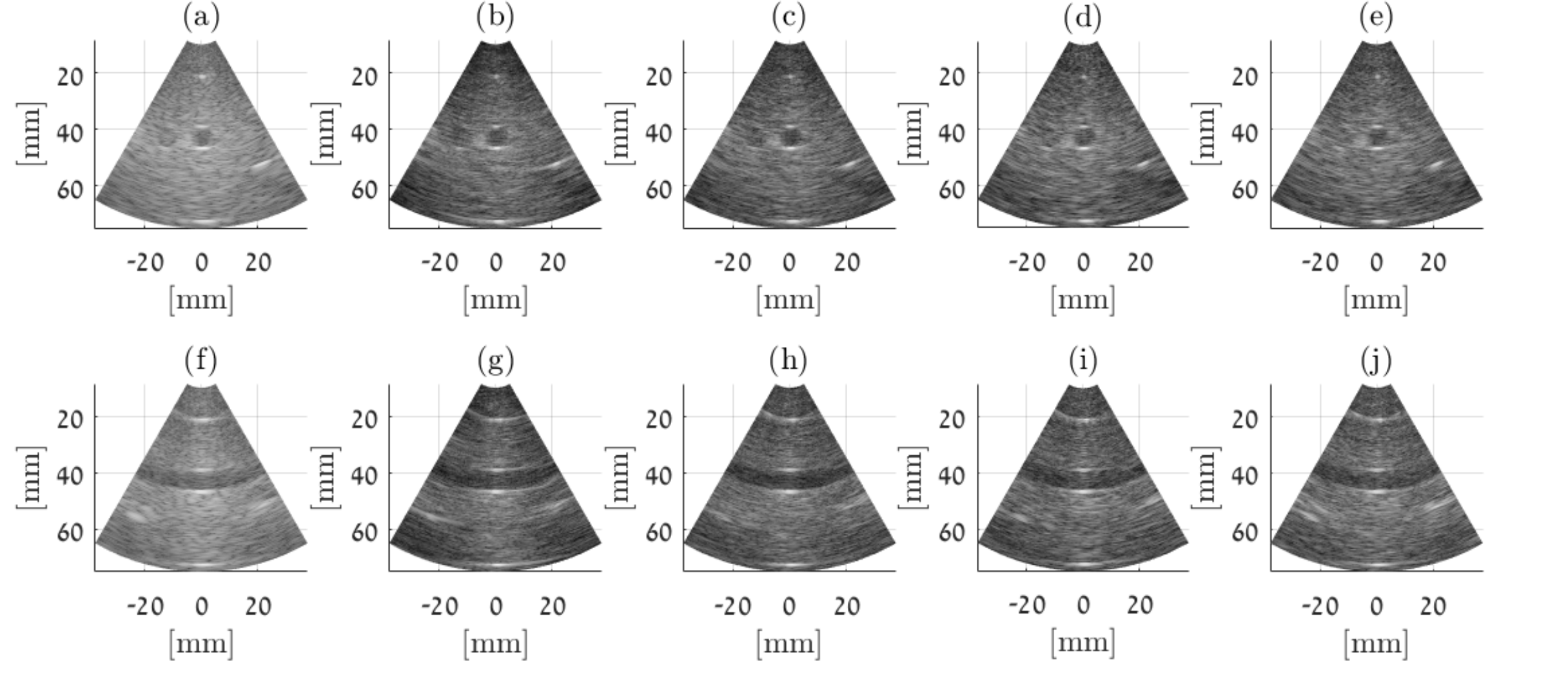}
 \caption{Orthogonal views of (top) $xz$ and (bottom) $yz$ planes  of a Gammex phantom with anechoic cyst obtained using diverging-waves with DAS (961), COBA (961), SCOBA I (225), SCOBA II (169), and SCOBA III (121) in their respective order from left to right. Numbers in brackets refer to the number of receive elements. Images are displayed with a dynamic range of 80dB.}
  \label{fig:dw_phantom_contrast}
 \end{figure*}
\begin{figure*}
\centering
\begin{subfigure}{.35\textwidth}
  \centering
  \includegraphics[trim={6cm 3cm 6cm 4cm},clip, height = 4cm, width = 0.8\linewidth]{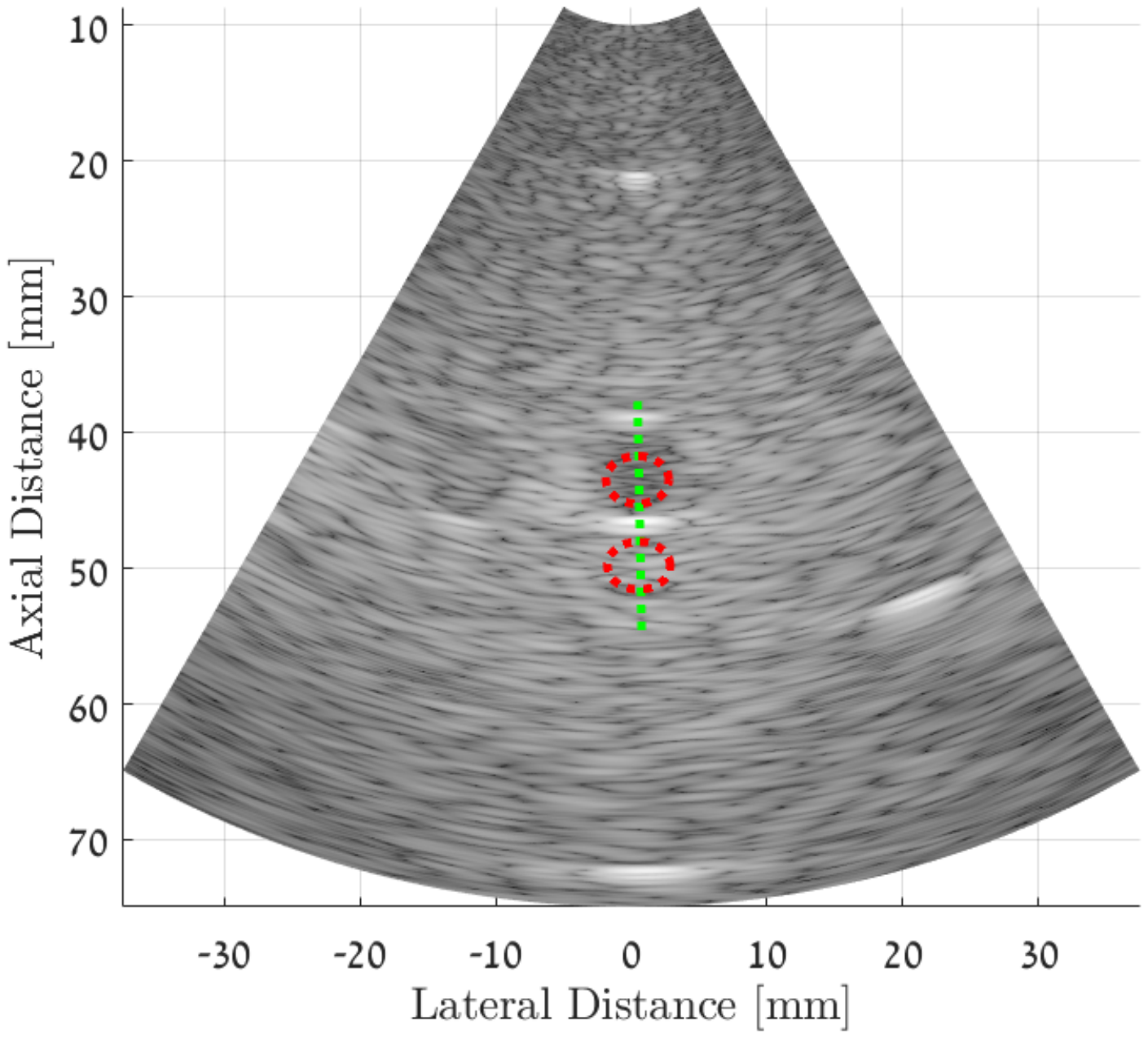}
  \caption{}
\end{subfigure}%
\begin{subfigure}{.55\textwidth}
  \centering
  \includegraphics[trim={2cm 3.5cm 1cm 4cm},clip, height = 4cm, width = 0.9\linewidth]{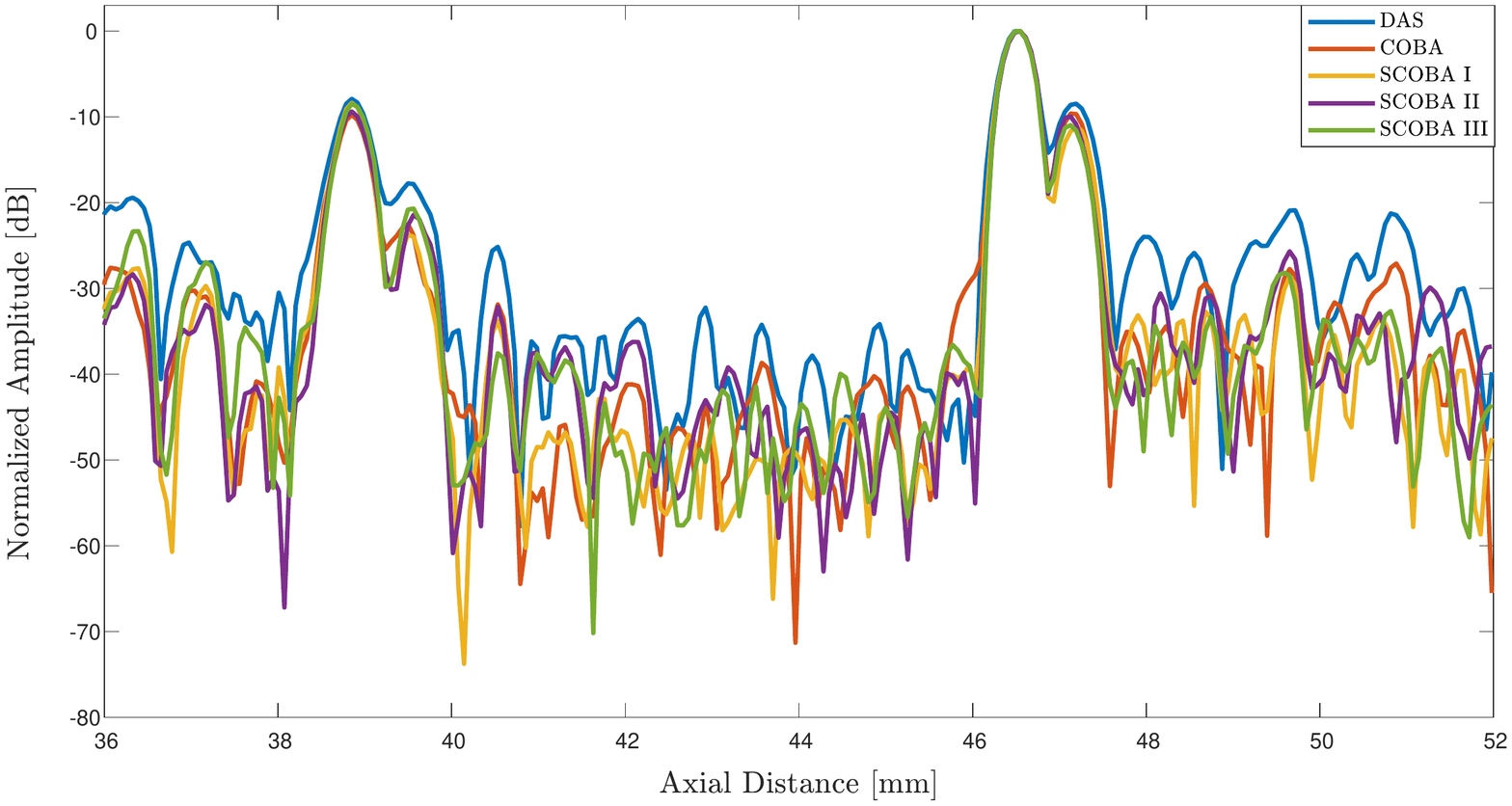}
  \caption{}
\end{subfigure}
\caption{(a) The red circles point to the areas used for computing the CR values of Table\,\ref{tab:contrast}, while the dashed green line marks the location of the cross-section shown in (b) to assess contrast.}
\label{fig:dw_contrast}
\end{figure*}

Next, we examine the proposed techniques when diverging-wave transmission is used to achieve high frame-rate. In Fig.\,\ref{fig:dw_phantom_contrast}, we present images of anechoic cyst, acquired with unfocused insonification and recovered by performing coherent compounding upon reception. These results along with the cross-sections shown in Fig.\,\ref{fig:dw_contrast} prove that the proposed techniques are suitable for diverging-wave transmission. COBA clearly demonstrates better contrast than DAS, while all variants of SCOBA outperform DAS in terms of contrast where their performance increases with the number of elements. The CR values given in Table\,\ref{tab:contrast} verify these conclusions. 

\begin{figure*}
 \centering
 \includegraphics[trim={3cm 4cm 2cm 6cm},clip,height = 7cm, width = 0.9\linewidth]{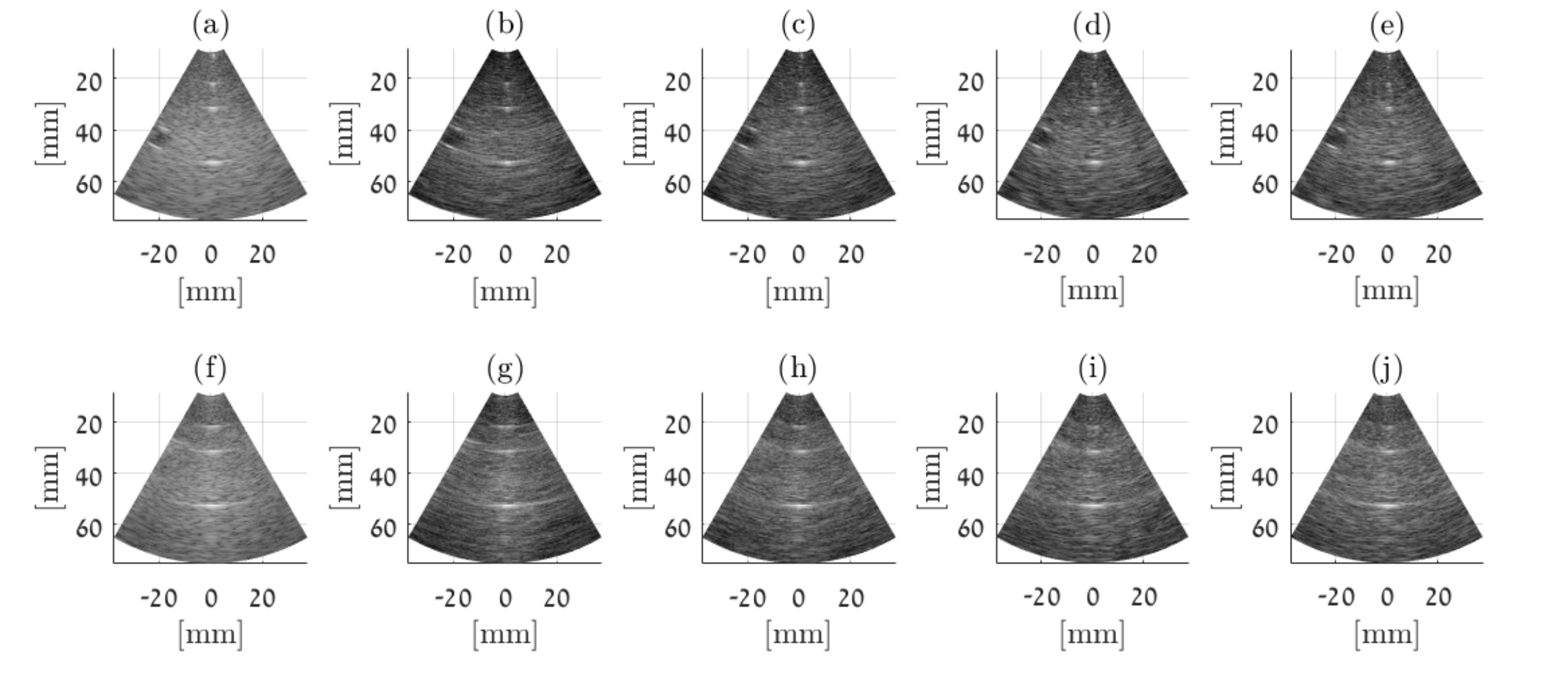}
 \caption{Orthogonal views of (top) $xz$ and (bottom) $yz$ planes  of a Gammex phantom with punctual targets obtained using diverging-waves with DAS (961), COBA (961), SCOBA I (225), SCOBA II (169), and SCOBA III (121) in their respective order from left to right. Numbers in brackets refer to the number of receive elements. Images are displayed with a dynamic range of 80dB.}
  \label{fig:dw_phantom_resolution}
 \end{figure*}

\begin{figure*}
\centering
\begin{subfigure}{.35\textwidth}
  \centering
  \includegraphics[trim={0cm 4cm 0cm 5cm},clip, height = 4cm, width = 0.7\linewidth]{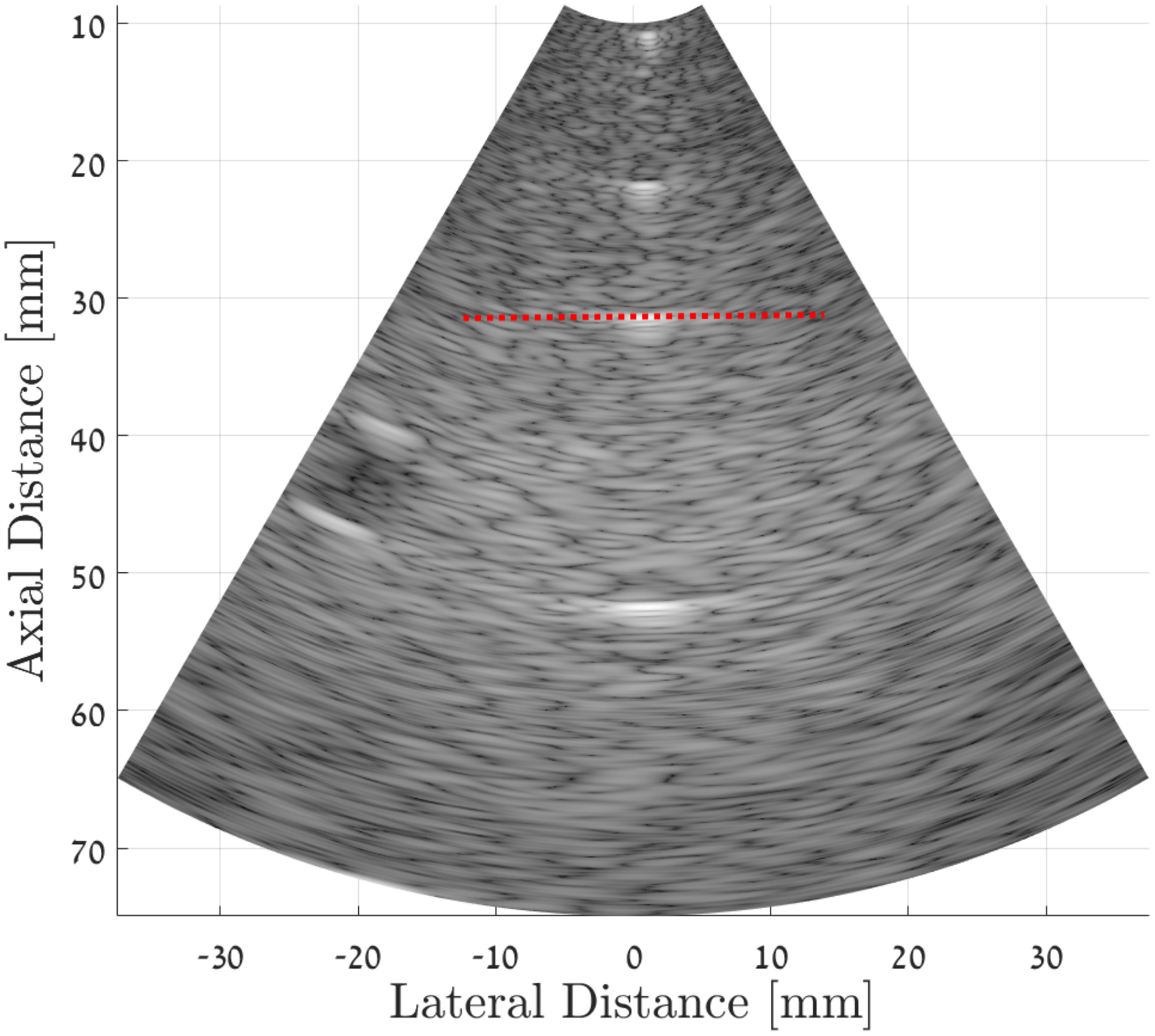}
  \caption{}
\end{subfigure}%
\begin{subfigure}{.55\textwidth}
  \centering
  \includegraphics[trim={2cm 3.5cm 1cm 4cm},clip, height = 4cm, width = 0.9\linewidth]{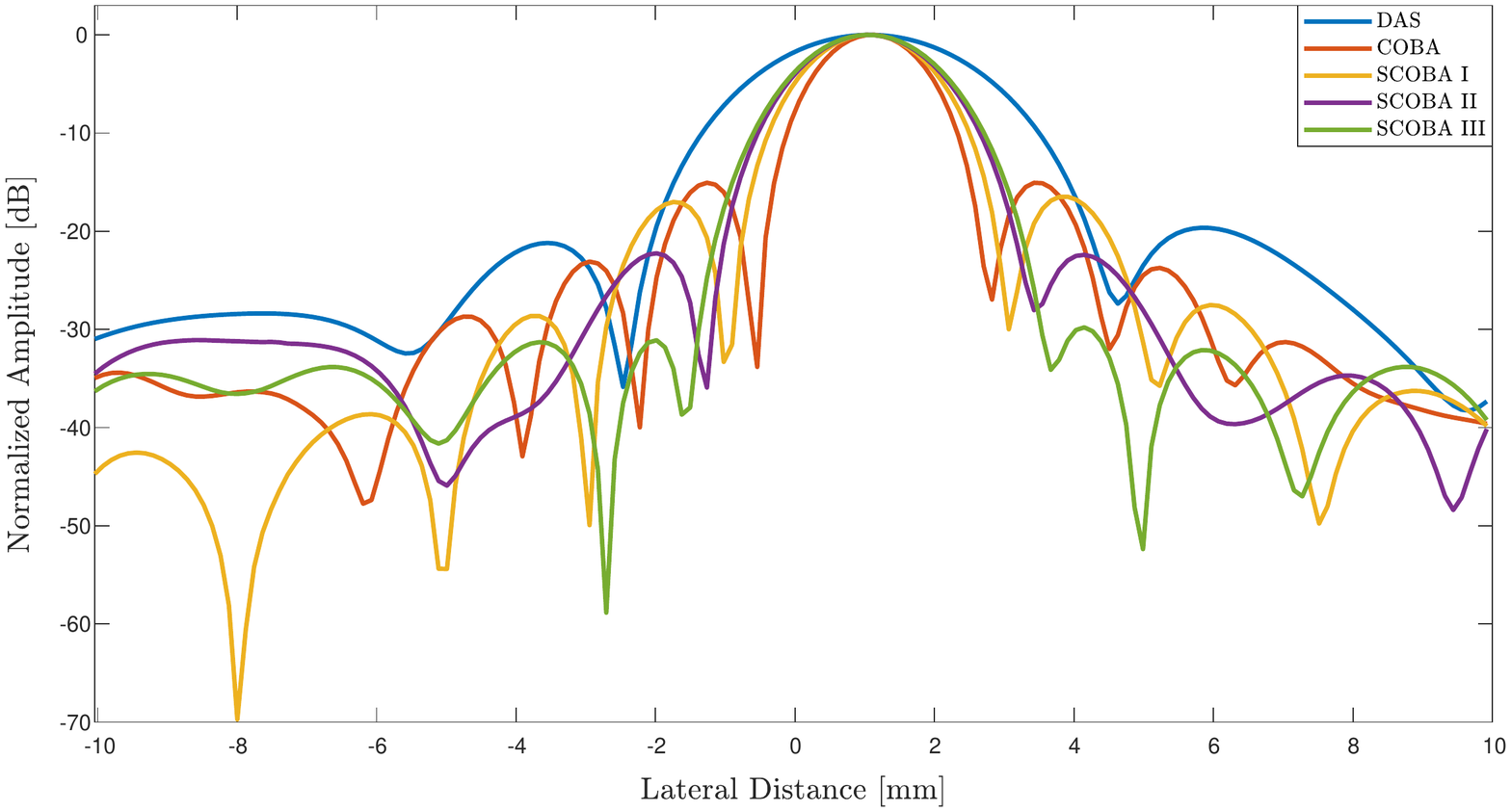}
  \caption{}
\end{subfigure}
\caption{(a) The dashed red line indicates the location of the cross-section shown in (b) to assess resolution.}
\label{fig:dw_resolution}
\end{figure*}

To estimate image resolution, we show additional images in Fig.\,\ref{fig:dw_phantom_resolution} which include point targets. The cross-section presented in Fig.\,\ref{fig:dw_resolution} displays the resolution improvement obtained by COBA and the selected versions of SCOBA. These results are supported quantitatively by the FWHM values in Table\,\ref{tab:resolution}. 

\begin{figure*}
\centering
\begin{subfigure}{.8\textwidth}
  \centering
  \includegraphics[trim={3cm 8cm 3cm 7cm},clip, height = 3cm, width = 1\linewidth]{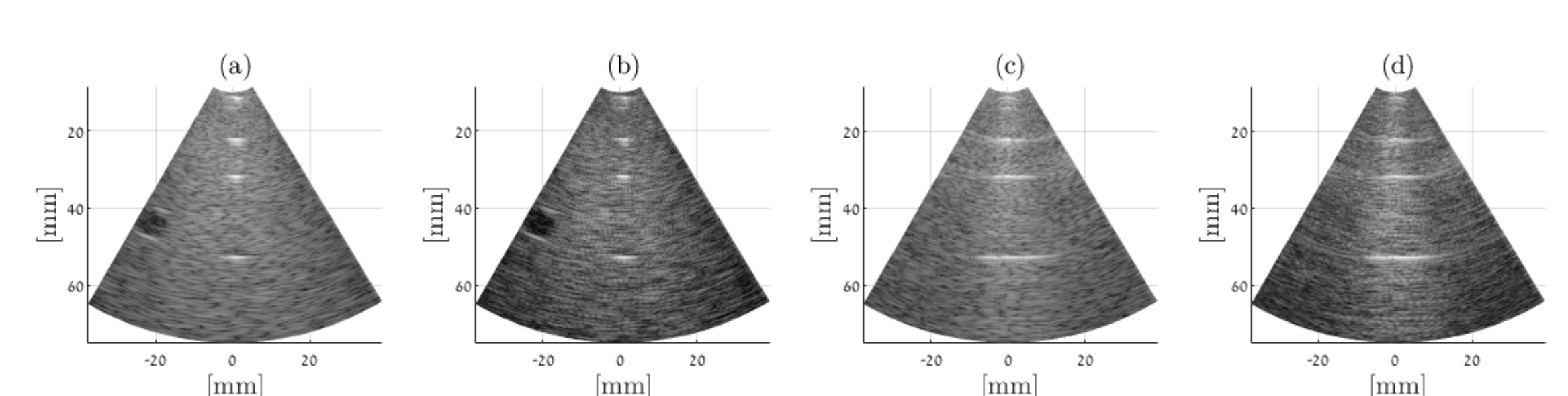}
\end{subfigure}%
\\
\begin{subfigure}{.8\textwidth}
  \centering
  \includegraphics[trim={3cm 8cm 3cm 7cm},clip, height = 3cm, width = 1\linewidth]{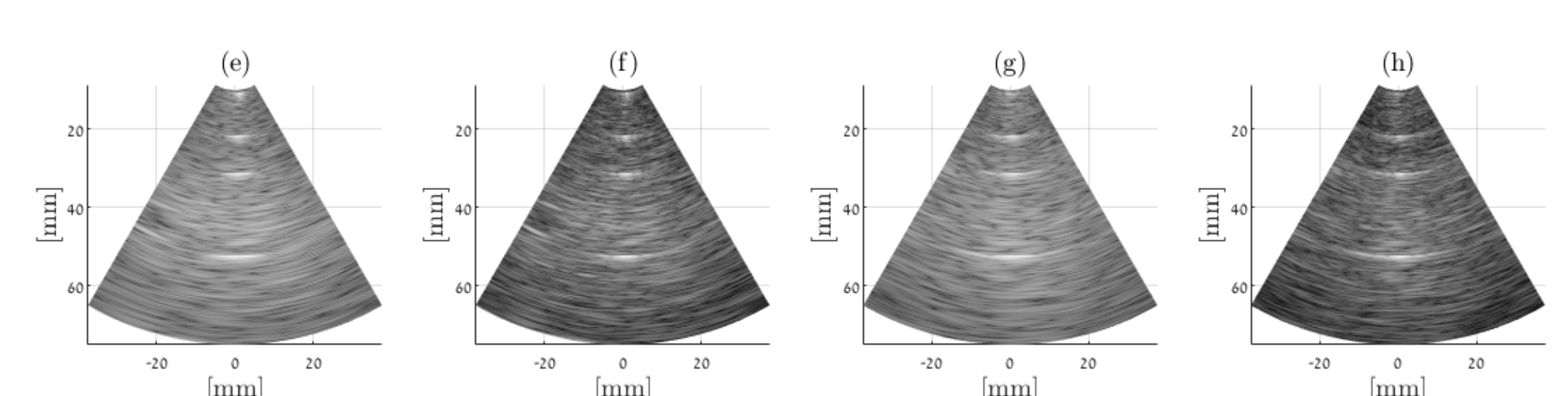}
\end{subfigure}%
\caption{Images of the $xz$ (left) and $yz$ (right) plane of a Gammex phantom acquired using a 161 ($13\times13)$ element UPA with either focused transmission (top) or diverging waves (bottom). Upon reception, we applied DAS on the full receive array to produce images (a), (c), (e) and (g). The other images, (b), (d), (f) and (h), were obtained by SCOBA operating on a 81 element fractal receive array shown in Fig.\,\ref{fig:fractals}(b). Images are displayed with a dynamic range of 80dB.}
\label{fig:phantom_fractal}
\end{figure*}

We complete this part by demonstrating the use of our recursive shceme where we expand a nine element UPA to create the fractal array $F_2$ shown in Fig.\,\ref{fig:fractals}(b).
The fractal sparse array is utilized to perform 3D imaging with either focused transmission or diverging-wave compounding where we apply SCOBA upon reception. Fig.\,\ref{fig:phantom_fractal} displays resultant images as well as images recovered by DAS operating on a $13\times13$ element array (full receive aperture). Assessing the images visually, one can see that use of the fractal array led to images that exhibit better resolution and contrast compared to those created by DAS, while utilizing fewer than half of the receive elements (81 out 169). These results show the simplicity and efficiency of the proposed fractal design and its  in performing sparse beamforming. Note that the images of Fig.\,\ref{fig:phantom_fractal} are of low quality compared to the previous results which is expected since here we use a considerably smaller aperture upon reception.  

\subsection{Ex-vivo Experiments}
\label{subsec:exvivo}
\begin{figure*}
 \centering
 \includegraphics[trim={4.5cm 8.5cm 3cm 8cm},clip,height = 3cm, width = 0.8\linewidth]{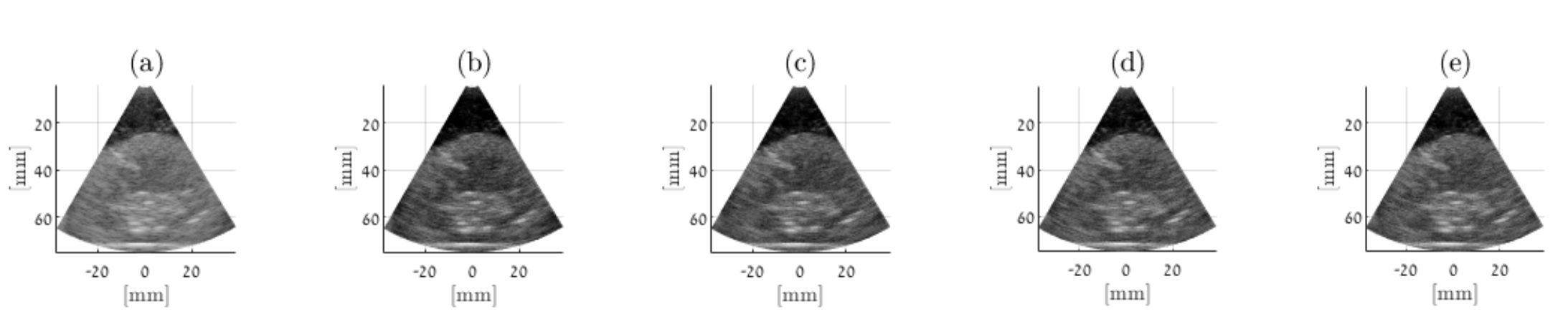}
 \caption{Images of the $xz$ plane of a lamb kidney using focused transmission with DAS (961), COBA (961), SCOBA I (225), SCOBA II (169), and SCOBA III (121) in their respective order from left to right. Numbers in brackets refer to the number of receive elements. Images are displayed with a dynamic range of 80dB.}
  \label{fig:exvivo_focus}
 \end{figure*}

\begin{figure*}[htb]
 \centering
 \includegraphics[trim={4.5cm 10cm 3cm 6.5cm},clip,height = 3cm, width = 0.8\linewidth]{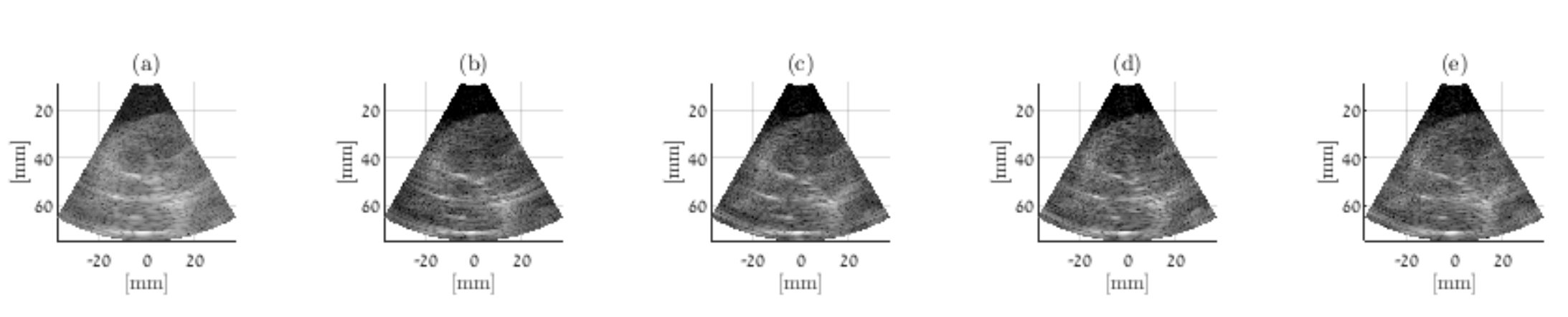}
 \caption{Images of the $xz$ plane of a lamb kidney obtained using diverging-waves with DAS (961), COBA (961), SCOBA I (225), SCOBA II (169), and SCOBA III (121) in their respective order from left to right. Numbers in brackets refer to the number of receive elements. Images are displayed with a dynamic range of 80dB.}
  \label{fig:exvivo_dw}
 \end{figure*}

Finally, \textit{ex vivo} acquisitions were performed, scanning a lamb kidney. This medium has a typical external shape and it also exhibits, in its internal structure, characteristics that should be found in 3D ultrasound imaging (e.g. vascularization). To properly assess the proposed beamforming strategy for \textit{ex vivo} acquisitions, we maintain the same transmission/reception settings as before. Given the size of the imaged tissues, the probe has been placed in order to scan the larger possible section of each medium. Fig.\,\ref{fig:exvivo_focus} presents various results acquired using focused transmission while Fig.\,\ref{fig:exvivo_dw} displays images obtained with diverging-wave compounding. 
It can be seen from the results that COBA and SCOBA yield images with higher contrast than DAS. Examining the images of Fig.\,\ref{fig:exvivo_dw}, we can clearly observe the improvement in resolution achieved by COBA and SCOBA. Thus, these images provide a strong evidence that the proposed techniques outperform DAS. SCOBA variants offer similar or better image quality than DAS while allowing a 4-8 fold element reduction, thus, enabling high-quality 3D ultrasound imaging.

\begin{figure*}
\centering
\begin{subfigure}{.4\textwidth}
  \centering
  \includegraphics[trim={3cm 4cm 3cm 3cm},clip, height = 4cm, width = 0.95\linewidth]{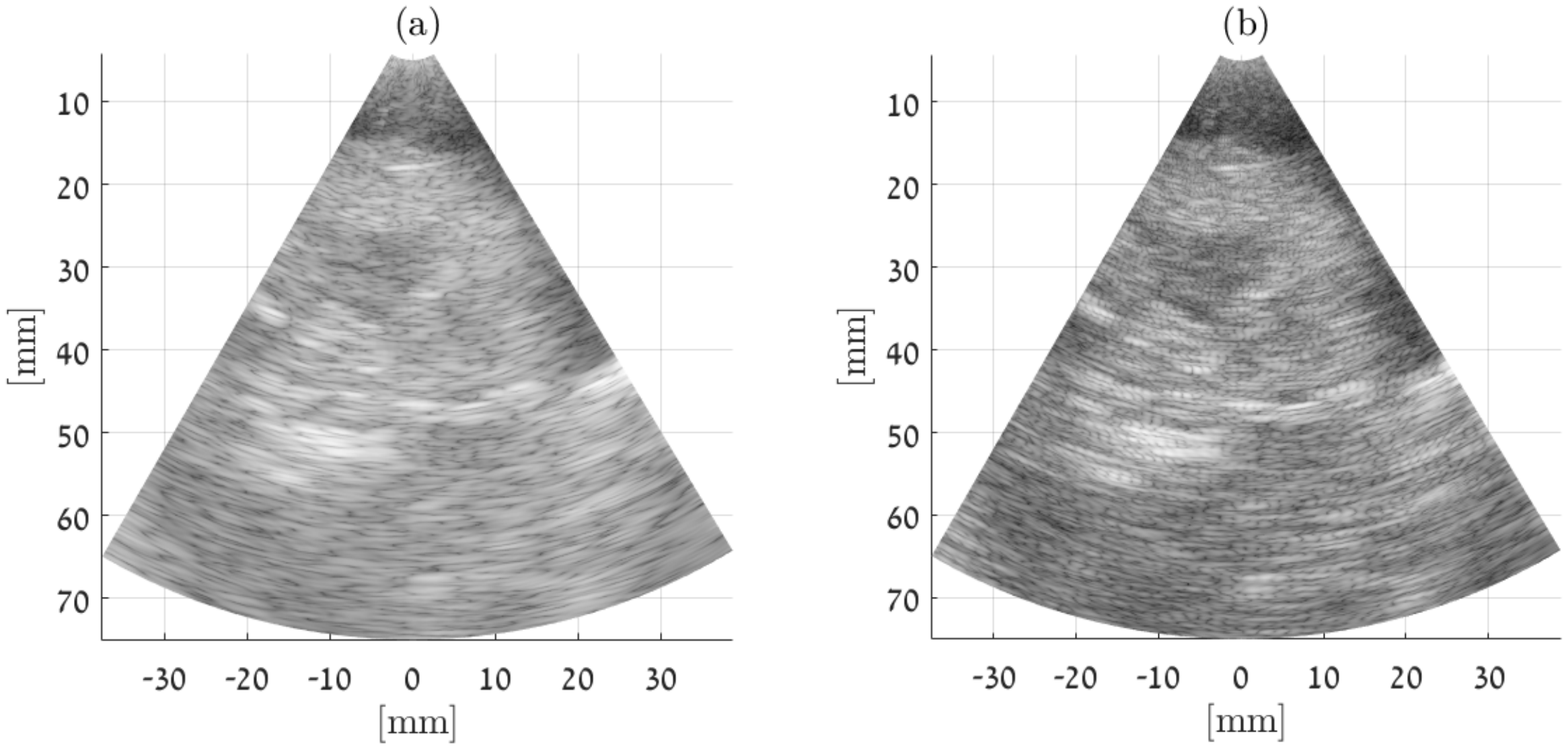}
\end{subfigure}%
\begin{subfigure}{.4\textwidth}
  \centering
  \includegraphics[trim={3cm 4cm 3cm 3cm},clip, height = 4cm, width = 0.95\linewidth]{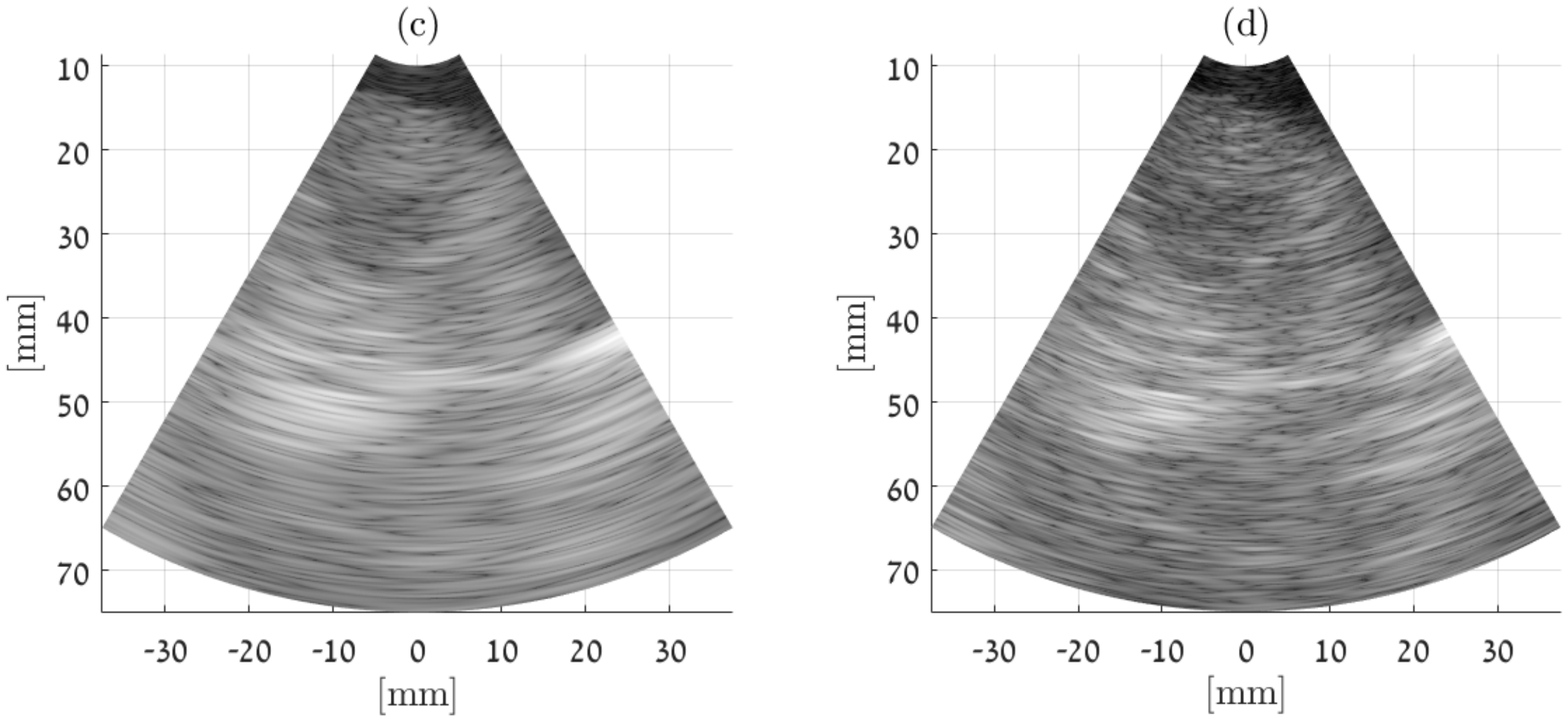}
\end{subfigure}
\caption{Images of the $xz$ plane of a lamb kidney. (a) and (c) are images obtained by DAS operating on a 161 ($13\times13)$ element receive UPA with focused and diverging wave transmission respectively. (b) and (d) are images obtained by SCOBA operating on a 81 element fractal receive array (Fig.\,\ref{fig:fractals}b) with focused and diverging wave transmission respectively.}
\label{fig:exvivo_fractal}
\end{figure*}

We end the experiments with last results obtained using the fractal array of Fig.\,\ref{fig:fractals}(b). The generated images are shown in Fig.\,\ref{fig:exvivo_fractal} in comparing to DAS as before. Theses results show the applicability of our recursive array design with SCOBA, leading to improved performance, superior to DAS, where we use significantly fewer elements than DAS without compromising image quality.

\section{Conclusion}
\label{sec:conclude}

In this work we introduced COBA-3D which extends the notion of convolutional beamforming to the 3D setting with unfocused insonification. The key part of COBA-3D is the 2D spatial convolution of the received signals, computed efficiently in the Fourier domain. This results in improved image resolution and contrast in comparison to DAS.
We relate this improvement in image quality to the virtual sum co-array which is larger than the physical array and thus yields an enhanced receive beam pattern. Furthermore, we presented SCOBA-3D which exploits sparse arrays whose sum co-arrays are large to perform 3D imaging with a notable 4-8 fold element reduction upon reception. To complete our approach, we introduced a fractal design which expands recursively a generator array with favorable properties to create an arbitrarily large sparse array with the same properties. This design facilities the construction of sparse arrays with multiple desired properties and its impact should increase as technology advances and the number of elements grows.      

To assess the performance of our proposed techniques, we performed various experiments on phantom scans, including focused and diverging-wave transmissions. The qualitative and quantitative results verify that COBA-3D achieves improved image resolution and contrast in comparison to DAS. In additions, SCOBA-3D enables to generate high-quality 3D images with a small number of receive elements, typically found in 1D probes. Similar results were obtained in \textit{ex-vivo} experiments, validating the proposed methods.

To summarize, convolutional beamforming for 3D imaging offers enhanced image quality in terms of both resolution and contrast. Moreover, it can be easily combined with unfocused insonification, such as diverging-wave compounding, to allow ultrafast frame rate. Finally, our fractal array design complements the proposed beamforming by allowing to construct sparse arrays where the majority of receive electronics are discarded. 
Thus, we reduce the processing rate, cost and power, facilitating the use of high-performance 3D US imaging with limited hardware.

\appendices
\section{Receive Beam Pattern Analysis}
\label{app:beampattern}

Here we derive an expression of the receive 2D beam pattern created by COBA-3D. The following can be seen as an extension to 2D of the beam pattern analysis given in \cite{cohen2018coba}. 

Assume the input signal is $f(t)=e^{j\omega_0t}$ impinging on the array at some direction $(\theta_0,\phi_0)$. Thus, we obtain
\begin{equation}
r_{n,m}(t)=e^{j\omega_0t}e^{-j\omega_0\tau_{n,m}},
\label{eq:sqrtsignal}
\end{equation} 
where $\tau_{n,m}$ is defined in (\ref{eq:delay}). 
The beamformed signal can be expressed as
\begin{equation}
    b(t)=\sum_{(u,v)\in E}\, \sum_{(k,l)\in E} (r_{u,v}\cdot r_{k,l})(t).
\end{equation}
Notice that
\begin{equation}
    (r_{u,v}\cdot r_{k,l})(t) = e^{j\omega_0t}r_{u+k,v+l}(t).
\end{equation}
Therefore, we get
\begin{equation}
    b(t)=\sum_{(n,m)\in S_E} \tilde{w}_R[n,m]a_E[n,m]e^{j\omega_0t}r_{n,m}(t),
    \label{eq:cobasignalwithsumcoarray}
\end{equation}
where $\tilde{w}_R[n,m]$ is the receive apodization weights, $S_E$ is the sum co-array of $E$ and $a_E[n,m]$ are the corresponding intrinsic apodization given by \eqref{eq:intrinsicapodizationentries}.
Substituting (\ref{eq:sqrtsignal}) into (\ref{eq:cobasignalwithsumcoarray}), we obtain 
\begin{equation}
    b(t) = e^{j2\omega_0t}\sum_{(n,m)\in S_E} w_R[n,m]e^{-j\omega_0\tau_{n,m}}.
\end{equation}
where we define $w_R[n,m]\triangleq \tilde{w}_R[n,m]a_E[n,m]$. Notice that the weights $\tilde{w}_R[n,m]$ should consider the intrinsic apodization to obtain a desired effective apodization $w_R[n,m]$. 

Consequently, the receive beam pattern of COBA-3D is
\begin{equation}
H_{COBA-3D}(\theta,\phi)=
\sum_{(n,m)\in S_E} w_R[n,m]e^{-j (s_xn+s_ym)}
\label{eq:cobaFourierbeampattern}
\end{equation}
where $s_x$ and $s_y$ are given by (\ref{eq:sptialfreqs}). The resultant beam pattern depends on the sum co-array and the intrinsic apodization. When the physical array $E$ is a UPA, then $S_E$ is another UPA of twice the aperture size in both axes. A larger aperture implies better image resolution. Moreover, it increases the degrees of freedom in determining the apodization. Hence, an appropriate choice of the weights, which accounts for the intrinsic apodization, can lead to effective weights such as Hamming apodization that enhance image contrast. 

\FloatBarrier
\bibliographystyle{IEEEtran}
\bibliography{IEEEabrv,REFS}

\begin{thebibliography}{10}
\providecommand{\url}[1]{#1}
\csname url@samestyle\endcsname
\providecommand{\newblock}{\relax}
\providecommand{\bibinfo}[2]{#2}
\providecommand{\BIBentrySTDinterwordspacing}{\spaceskip=0pt\relax}
\providecommand{\BIBentryALTinterwordstretchfactor}{4}
\providecommand{\BIBentryALTinterwordspacing}{\spaceskip=\fontdimen2\font plus
\BIBentryALTinterwordstretchfactor\fontdimen3\font minus
  \fontdimen4\font\relax}
\providecommand{\BIBforeignlanguage}[2]{{%
\expandafter\ifx\csname l@#1\endcsname\relax
\typeout{** WARNING: IEEEtran.bst: No hyphenation pattern has been}%
\typeout{** loaded for the language `#1'. Using the pattern for}%
\typeout{** the default language instead.}%
\else
\language=\csname l@#1\endcsname
\fi
#2}}
\providecommand{\BIBdecl}{\relax}
\BIBdecl

\bibitem{fenster2015ultrasound}
A.~Fenster and J.~C. Lacefield, \emph{Ultrasound imaging and therapy}.\hskip
  1em plus 0.5em minus 0.4em\relax Taylor \& Francis, 2015.

\bibitem{thomenius1996evolution}
K.~E. Thomenius, ``Evolution of ultrasound beamformers,'' in \emph{Proceedings
  of Ultrasonics Symposium.}, vol.~2.\hskip 1em plus 0.5em minus 0.4em\relax
  IEEE, 1996, pp. 1615--1622.

\bibitem{karaman1995synthetic}
M.~Karaman, P.-C. Li, and M.~O'Donnell, ``Synthetic aperture imaging for small
  scale systems,'' \emph{IEEE Transactions on Ultrasonics, Ferroelectrics, and
  Frequency Control}, vol.~42, no.~3, pp. 429--442, 1995.

\bibitem{eklund2013medical}
A.~Eklund, P.~Dufort, D.~Forsberg, and S.~M. LaConte, ``Medical image
  processing on the {GPU}: Past, present and future,'' \emph{Medical Image
  Analysis}, vol.~17, no.~8, pp. 1073--1094, 2013.

\bibitem{petrusca2018fast}
L.~Petrusca, F.~Varray, R.~Souchon, A.~Bernard, J.-Y. Chapelon, H.~Liebgott,
  W.~N’Djin, and M.~Viallon, ``Fast volumetric ultrasound {B}-mode and
  {D}oppler imaging with a new high-channels density platform for advanced 4{D}
  cardiac imaging/therapy,'' \emph{Applied Sciences}, vol.~8, no.~2, p. 200,
  2018.

\bibitem{jensen2013sarus}
J.~A. Jensen, H.~Holten-Lund, R.~T. Nilsson, M.~Hansen, U.~D. Larsen, R.~P.
  Domsten, B.~G. Tomov, M.~B. Stuart, S.~I. Nikolov, M.~J. Pihl \emph{et~al.},
  ``{SARUS}: A synthetic aperture real-time ultrasound system,'' \emph{IEEE
  Transactions on Ultrasonics, Ferroelectrics, and Frequency Control}, vol.~60,
  no.~9, pp. 1838--1852, 2013.

\bibitem{provost20153}
J.~Provost, C.~Papadacci, C.~Demene, J.-L. Gennisson, M.~Tanter, and M.~Pernot,
  ``3-{D} ultrafast {D}oppler imaging applied to the noninvasive mapping of
  blood vessels in {V}ivo,'' \emph{IEEE Transactions on Ultrasonics,
  Ferroelectrics, and Frequency Control}, vol.~62, no.~8, pp. 1467--1472, 2015.

\bibitem{savord2003fully}
B.~Savord and R.~Solomon, ``Fully sampled matrix transducer for real time 3{D}
  ultrasonic imaging,'' in \emph{Symposium on Ultrasonics.}, vol.~1.\hskip 1em
  plus 0.5em minus 0.4em\relax IEEE, 2003, pp. 945--953.

\bibitem{wildes20164}
D.~Wildes, W.~Lee, B.~Haider, S.~Cogan, K.~Sundaresan, D.~M. Mills, C.~Yetter,
  P.~H. Hart, C.~R. Haun, M.~Concepcion \emph{et~al.}, ``{4D ICE}: A {2D} array
  transducer with integrated {ASIC} in a 10-{Fr} catheter for real-time {3D}
  intracardiac echocardiography,'' \emph{IEEE Transactions on Ultrasonics,
  Ferroelectrics, and Frequency Control}, vol.~63, no.~12, pp. 2159--2173,
  2016.

\bibitem{santos2016diverging}
P.~Santos, G.~U. Haugen, L.~L{\o}vstakken, E.~Samset, and J.~D’hooge,
  ``Diverging wave volumetric imaging using subaperture beamforming,''
  \emph{IEEE Transactions on Ultrasonics, Ferroelectrics, and Frequency
  Control}, vol.~63, no.~12, pp. 2114--2124, 2016.

\bibitem{matrone2014volumetric}
G.~Matrone, A.~S. Savoia, M.~Terenzi, G.~Caliano, F.~Quaglia, and G.~Magenes,
  ``A volumetric {CMUT}-based ultrasound imaging system simulator with
  integrated reception and $\mu$-beamforming electronics models,'' \emph{IEEE
  Transactions on Ultrasonics, Ferroelectrics, and Frequency Control}, vol.~61,
  no.~5, pp. 792--804, 2014.

\bibitem{bhuyan2013integrated}
A.~Bhuyan, J.~W. Choe, B.~C. Lee, I.~O. Wygant, A.~Nikoozadeh, {\"O}.~Oralkan,
  and B.~T. Khuri-Yakub, ``Integrated circuits for volumetric ultrasound
  imaging with {2D CMUT} arrays,'' \emph{IEEE Transactions on Biomedical
  Circuits and Systems}, vol.~7, no.~6, pp. 796--804, 2013.

\bibitem{kortbek2013sequential}
J.~Kortbek, J.~A. Jensen, and K.~L. Gammelmark, ``Sequential beamforming for
  synthetic aperture imaging,'' \emph{Ultrasonics}, vol.~53, no.~1, pp. 1--16,
  2013.

\bibitem{fisher2005reconfigurable}
R.~Fisher, K.~Thomenius, R.~Wodnicki, R.~Thomas, S.~Cogan, C.~Hazard, W.~Lee,
  D.~Mills, B.~Khuri-Yakub, A.~Ergun \emph{et~al.}, ``Reconfigurable arrays for
  portable ultrasound,'' in \emph{IEEE Ultrasonics Symposium}, vol.~1, 2005,
  pp. 495--499.

\bibitem{wygant2009integrated}
I.~O. Wygant, N.~S. Jamal, H.~J. Lee, A.~Nikoozadeh, {\"O}.~Oralkan,
  M.~Karaman, and B.~T. Khuri-Yakub, ``An integrated circuit with transmit
  beamforming flip-chip bonded to a {2D CMUT} array for {3D} ultrasound
  imaging,'' \emph{IEEE Transactions on Ultrasonics, Ferroelectrics, and
  Frequency Control}, vol.~56, no.~10, pp. 2145--2156, 2009.

\bibitem{smith1991high}
S.~W. Smith, H.~G. Pavy, and O.~T. von Ramm, ``High-speed ultrasound volumetric
  imaging system. {I}. {T}ransducer design and beam steering,'' \emph{IEEE
  Transactions on Ultrasonics, Ferroelectrics, and Frequency Control}, vol.~38,
  no.~2, pp. 100--108, 1991.

\bibitem{von1991high}
O.~T. Von~Ramm, S.~W. Smith, and H.~G. Pavy, ``High-speed ultrasound volumetric
  imaging system. {II. P}arallel processing and image display,'' \emph{IEEE
  Transactions on Ultrasonics, Ferroelectrics, and Frequency Control}, vol.~38,
  no.~2, pp. 109--115, 1991.

\bibitem{lok2018microbeamforming}
U.-W. Lok and P.-C. Li, ``Microbeamforming with error compensation,''
  \emph{IEEE Transactions on Ultrasonics, Ferroelectrics, and Frequency
  Control}, vol.~65, no.~7, pp. 1153--1165, 2018.

\bibitem{fuller2005sonic}
M.~I. Fuller, E.~V. Brush, M.~D. Eames, T.~N. Blalock, J.~A. Hossack, and W.~F.
  Walker, ``The sonic window: second generation prototype of low-cost,
  fully-integrated, pocket-sized medical ultrasound device,'' in \emph{IEEE
  Ultrasonics Symposium.}, vol.~1, 2005, pp. 273--276.

\bibitem{lee2004miniaturized}
W.~Lee, S.~F. Idriss, P.~D. Wolf, and S.~W. Smith, ``A miniaturized catheter
  2-{D} array for real-time, 3-{D} intracardiac echocardiography,'' \emph{IEEE
  Transactions on Ultrasonics, Ferroelectrics, and Frequency Control}, vol.~51,
  no.~10, pp. 1334--1346, 2004.

\bibitem{chen2011cmut}
A.~I. Chen, L.~L. Wong, A.~S. Logan, and J.~T. Yeow, ``A {CMUT}-based real-time
  volumetric ultrasound imaging system with row-column addressing,'' in
  \emph{IEEE International Ultrasonics Symposium (IUS)}, 2011, pp. 1755--1758.

\bibitem{rasmussen20133d}
M.~F. Rasmussen and J.~A. Jensen, ``3{D} ultrasound imaging performance of a
  row-column addressed 2{D} array transducer: A simulation study,'' in
  \emph{Medical Imaging 2013: Ultrasonic Imaging, Tomography, and Therapy},
  vol. 8675, 2013, p. 86750C.

\bibitem{rasmussen20133}
------, ``3{D} ultrasound imaging performance of a row-column addressed 2{D}
  array transducer: A measurement study,'' in \emph{International Ultrasonics
  Symposium (IUS),}.\hskip 1em plus 0.5em minus 0.4em\relax IEEE, 2013, pp.
  1460--1463.

\bibitem{rasmussen20153}
M.~F. Rasmussen, T.~L. Christiansen, E.~V. Thomsen, and J.~A. Jensen, ``3{D}
  imaging using row-column-addressed arrays with integrated apodization -
  {P}art {I}: Apodization design and line element beamforming,'' \emph{IEEE
  Transactions on Ultrasonics, Ferroelectrics, and Frequency Control}, vol.~62,
  no.~5, pp. 947--958, 2015.

\bibitem{christiansen20153}
T.~L. Christiansen, M.~F. Rasmussen, J.~P. Bagge, L.~N. Moesner, J.~A. Jensen,
  and E.~V. Thomsen, ``3{D} imaging using row--column-addressed arrays with
  integrated apodization — {P}art {II}: Transducer fabrication and
  experimental results,'' \emph{IEEE Transactions on Ultrasonics,
  Ferroelectrics, and Frequency Control}, vol.~62, no.~5, pp. 959--971, 2015.

\bibitem{daya2017compensated}
I.~B. Daya, A.~I. Chen, M.~J. Shafiee, A.~Wong, and J.~T. Yeow, ``Compensated
  row-column ultrasound imaging system using multilayered edge guided
  stochastically fully connected random fields,'' \emph{Scientific reports},
  vol.~7, no.~1, p. 10644, 2017.

\bibitem{bouzari2017curvilinear}
H.~Bouzari, M.~Engholm, C.~Beers, M.~B. Stuart, S.~I. Nikolov, E.~V. Thomsen,
  and J.~A. Jensen, ``Curvilinear 3{D} imaging using row-column-addressed 2{D}
  arrays with a diverging lens: Feasibility study,'' \emph{IEEE Transactions on
  Ultrasonics, Ferroelectrics, and Frequency Control}, vol.~64, no.~6, pp.
  978--988, 2017.

\bibitem{flesch20174d}
M.~Flesch, M.~Pernot, J.~Provost, G.~Ferin, A.~Nguyen-Dinh, M.~Tanter, and
  T.~Deffieux, ``4{D} in vivo ultrafast ultrasound imaging using a row-column
  addressed matrix and coherently-compounded orthogonal plane waves,''
  \emph{Physics in Medicine \& Biology}, vol.~62, no.~11, p. 4571, 2017.

\bibitem{logan201132}
A.~S. Logan, L.~L. Wong, A.~I. Chen, and J.~T. Yeow, ``A 32 x 32 element
  row-column addressed capacitive micromachined ultrasonic transducer,''
  \emph{IEEE Transactions on Ultrasonics, Ferroelectrics, and Frequency
  Control}, vol.~58, no.~6, pp. 1266--1271, 2011.

\bibitem{savoia2007p2b}
A.~Savoia, V.~Bavaro, G.~Caliano, A.~Caronti, R.~Carotenuto, P.~Gatta,
  C.~Longo, and M.~Pappalardo, ``P2{B}-4 crisscross 2{D cMUT} array:
  Beamforming strategy and synthetic 3{D} imaging results,'' in \emph{IEEE
  Ultrasonics Symposium Proceedings}, 2007, pp. 1514--1517.

\bibitem{yang2014separable}
M.~Yang, R.~Sampson, S.~Wei, T.~F. Wenisch, and C.~Chakrabarti, ``Separable
  beamforming for {3D} medical ultrasound imaging,'' \emph{IEEE Transactions on
  Signal Processing}, vol.~63, no.~2, pp. 279--290, 2014.

\bibitem{owen2012application}
K.~Owen, M.~I. Fuller, and J.~A. Hossack, ``Application of {XY} separable {2D}
  array beamforming for increased frame rate and energy efficiency in handheld
  devices,'' \emph{IEEE Transactions on Ultrasonics, Ferroelectrics, and
  Frequency Control}, vol.~59, no.~7, pp. 1332--1343, 2012.

\bibitem{nikolov2003investigation}
S.~I. Nikolov and J.~A. Jensen, ``Investigation of the feasibility of {3D}
  synthetic aperture imaging,'' in \emph{IEEE Symposium on Ultrasonics},
  vol.~2, 2003, pp. 1903--1906.

\bibitem{jensen2006synthetic}
J.~A. Jensen, S.~I. Nikolov, K.~L. Gammelmark, and M.~H. Pedersen, ``Synthetic
  aperture ultrasound imaging,'' \emph{Ultrasonics}, vol.~44, pp. e5--e15,
  2006.

\bibitem{kortbek2008synthetic}
J.~Kortbek, J.~A. Jensen, and K.~L. Gammelmark, ``Synthetic aperture sequential
  beamforming,'' in \emph{IEEE Ultrasonics Symposium}, 2008, pp. 966--969.

\bibitem{wygant2006beamforming}
I.~O. Wygant, M.~Karaman, {\"O}.~Oralkan, and B.~T. Khuri-Yakub, ``Beamforming
  and hardware design for a multichannel front-end integrated circuit for
  real-time {3D} catheter-based ultrasonic imaging,'' in \emph{Medical Imaging:
  Ultrasonic Imaging and Signal Processing}, vol. 6147.\hskip 1em plus 0.5em
  minus 0.4em\relax International Society for Optics and Photonics, 2006, p.
  61470A.

\bibitem{lokesh2019diverging}
B.~Lokesh and A.~K. Thittai, ``Diverging beam transmit through limited
  aperture: A method to reduce ultrasound system complexity and yet obtain
  better image quality at higher frame rates,'' \emph{Ultrasonics}, vol.~91,
  pp. 150--160, 2019.

\bibitem{bottenus2013synthetic}
N.~Bottenus, B.~C. Byram, J.~J. Dahl, and G.~E. Trahey, ``Synthetic aperture
  focusing for short-lag spatial coherence imaging,'' \emph{IEEE Transactions
  on Ultrasonics, Ferroelectrics, and Frequency Control}, vol.~60, no.~9, pp.
  1816--1826, 2013.

\bibitem{eldar2012compressed}
Y.~C. Eldar and G.~Kutyniok, \emph{Compressed sensing: Theory and
  applications}.\hskip 1em plus 0.5em minus 0.4em\relax Cambridge University
  Press, 2012.

\bibitem{eldar2015sampling}
Y.~C. Eldar, \emph{Sampling theory: Beyond bandlimited systems}.\hskip 1em plus
  0.5em minus 0.4em\relax Cambridge University Press, 2015.

\bibitem{tur2011innovation}
R.~Tur, Y.~C. Eldar, and Z.~Friedman, ``Innovation rate sampling of pulse
  streams with application to ultrasound imaging,'' \emph{IEEE Transactions on
  Signal Processing}, vol.~59, no.~4, pp. 1827--1842, 2011.

\bibitem{zhuang2012ultrasonic}
X.~Zhuang, Y.~Zhao, Z.~Dai, H.~Wang, and L.~Wang, ``Ultrasonic signal
  compressive detection with sub-{N}yquist sampling rate,'' \emph{Journal of
  scientific and industrial research}, 2012.

\bibitem{zhou2013compressed}
J.~Zhou, Y.~He, M.~Chirala, B.~M. Sadler, and S.~Hoyos, ``Compressed digital
  beamformer with asynchronous sampling for ultrasound imaging,'' in \emph{IEEE
  International Conference on Acoustics, Speech and Signal Processing}, 2013,
  pp. 1056--1060.

\bibitem{liebgott2013pre}
H.~Liebgott, R.~Prost, and D.~Friboulet, ``Pre-beamformed {RF} signal
  reconstruction in medical ultrasound using compressive sensing,''
  \emph{Ultrasonics}, vol.~53, no.~2, pp. 525--533, 2013.

\bibitem{achim2010compressive}
A.~Achim, B.~Buxton, G.~Tzagkarakis, and P.~Tsakalides, ``Compressive sensing
  for ultrasound {RF} echoes using a-stable distributions,'' in \emph{Annual
  International Conference of the IEEE Engineering in Medicine and
  Biology}.\hskip 1em plus 0.5em minus 0.4em\relax IEEE, 2010, pp. 4304--4307.

\bibitem{tzagkarakis2013joint}
G.~Tzagkarakis, A.~Achim, P.~Tsakalides, and J.-L. Starck, ``Joint
  reconstruction of compressively sensed ultrasound {RF} echoes by exploiting
  temporal correlations,'' in \emph{IEEE 10th International Symposium on
  Biomedical Imaging}, 2013, pp. 632--635.

\bibitem{quinsac2012frequency}
C.~Quinsac, A.~Basarab, and D.~Kouam{\'e}, ``Frequency domain compressive
  sampling for ultrasound imaging,'' \emph{Advances in Acoustics and
  Vibration}, vol. 2012, 2012.

\bibitem{wagner2012compressed}
N.~Wagner, Y.~C. Eldar, and Z.~Friedman, ``Compressed beamforming in ultrasound
  imaging,'' \emph{IEEE Transactions on Signal Processing}, vol.~60, no.~9, pp.
  4643--4657, 2012.

\bibitem{chernyakova2014fourier}
T.~Chernyakova and Y.~C. Eldar, ``Fourier-domain beamforming: the path to
  compressed ultrasound imaging,'' \emph{IEEE Transactions on Ultrasonics,
  Ferroelectrics, and Frequency Control}, vol.~61, no.~8, pp. 1252--1267, 2014.

\bibitem{gedalyahu2011multichannel}
K.~Gedalyahu, R.~Tur, and Y.~C. Eldar, ``Multichannel sampling of pulse streams
  at the rate of innovation,'' \emph{IEEE Transactions on Signal Processing},
  vol.~59, no.~4, pp. 1491--1504, 2011.

\bibitem{baransky2014sub}
E.~Baransky, G.~Itzhak, N.~Wagner, I.~Shmuel, E.~Shoshan, and Y.~Eldar,
  ``Sub-{N}yquist radar prototype: Hardware and algorithm,'' \emph{IEEE
  Transactions on Aerospace and Electronic Systems}, vol.~50, no.~2, pp.
  809--822, 2014.

\bibitem{chernyakova2018fourier}
T.~Chernyakova, R.~Cohen, R.~Mulayoff, Y.~Sde-Chen, C.~Fraschini, J.~Bercoff,
  and Y.~C. Eldar, ``Fourier domain beamforming and structure-based
  reconstruction for plane-wave imaging,'' \emph{Transactions on Ultrasonics,
  Ferroelectrics, and Frequency Control}, vol.~65, no.~10, pp. 1810--1821,
  2018.

\bibitem{burshtein2016sub}
A.~Burshtein, M.~Birk, T.~Chernyakova, A.~Eilam, A.~Kempinski, and Y.~C. Eldar,
  ``Sub-{N}yquist sampling and {F}ourier domain beamforming in volumetric
  ultrasound imaging,'' \emph{IEEE Trans. Ultrason., Ferroelectr., Freq.
  Control}, vol.~63, no.~5, pp. 703--716, 2016.

\bibitem{liu2017compressed}
J.~Liu, Q.~He, and J.~Luo, ``A compressed sensing strategy for synthetic
  transmit aperture ultrasound imaging,'' \emph{IEEE Transactions on Medical
  Imaging}, vol.~36, no.~4, pp. 878--891, 2017.

\bibitem{davidsen1994two}
R.~E. Davidsen, J.~A. Jensen, and S.~W. Smith, ``Two-dimensional random arrays
  for real time volumetric imaging,'' \emph{Ultrasonic Imaging}, vol.~16,
  no.~3, pp. 143--163, 1994.

\bibitem{brunke1997broad}
S.~S. Brunke and G.~R. Lockwood, ``Broad-bandwidth radiation patterns of sparse
  two-dimensional vernier arrays,'' \emph{IEEE Transactions on Ultrasonics,
  Ferroelectrics, and Frequency Control}, vol.~44, no.~5, pp. 1101--1109, 1997.

\bibitem{yen2000sparse}
J.~T. Yen, J.~P. Steinberg, and S.~W. Smith, ``Sparse 2{D} array design for
  real time rectilinear volumetric imaging,'' \emph{IEEE Transactions on
  Ultrasonics, Ferroelectrics, and Frequency Control}, vol.~47, no.~1, pp.
  93--110, 2000.

\bibitem{austeng2002sparse}
A.~Austeng and S.~Holm, ``Sparse 2{D} arrays for 3{D} phased array
  imaging-design methods,'' \emph{IEEE Transactions on Ultrasonics,
  Ferroelectrics, and Frequency Control}, vol.~49, no.~8, pp. 1073--1086, 2002.

\bibitem{karaman2009minimally}
M.~Karaman, I.~O. Wygant, {\"O}.~Oralkan, and B.~T. Khuri-Yakub, ``Minimally
  redundant 2{D} array designs for 3{D} medical ultrasound imaging,''
  \emph{IEEE Transactions on Medical Imaging}, vol.~28, no.~7, pp. 1051--1061,
  2009.

\bibitem{diarra2013design}
B.~Diarra, M.~Robini, P.~Tortoli, C.~Cachard, and H.~Liebgott, ``Design of
  optimal 2{D} nongrid sparse arrays for medical ultrasound,'' \emph{IEEE
  Transactions on Biomedical Engineering}, vol.~60, no.~11, pp. 3093--3102,
  2013.

\bibitem{ramadas2014application}
S.~Ramadas, J.~Jackson, J.~Dziewierz, R.~O'Leary, and A.~Gachagan,
  ``Application of conformal map theory for design of 2{D} ultrasonic array
  structure for {NDT} imaging application: A feasibility study,'' \emph{IEEE
  Transactions on Ultrasonics, Ferroelectrics, and Frequency Control}, vol.~61,
  no.~3, pp. 496--504, 2014.

\bibitem{emmanuel2017validation}
E.~Roux, E.~Badescu, L.~Petrusca, F.~Varray, A.~Ramalli, C.~Cachard, M.~C.
  Robini, H.~Liebgott, and P.~Tortoli, ``Validation of optimal 2{D} sparse
  arrays in focused mode: Phantom experiments,'' in \emph{IEEE International
  Ultrasonic Symposium (IUS)}, 2017.

\bibitem{mitra2010general}
S.~K. Mitra, K.~Mondal, M.~K. Tchobanou, and G.~J. Dolecek, ``General
  polynomial factorization-based design of sparse periodic linear arrays,''
  \emph{IEEE Transactions on Ultrasonics, Ferroelectrics, and Frequency
  Fontrol}, vol.~57, no.~9, pp. 1952--1966, 2010.

\bibitem{roux2016wideband}
E.~Roux, A.~Ramalli, H.~Liebgott, C.~Cachard, M.~C. Robini, and P.~Tortoli,
  ``Wideband 2{D} array design optimization with fabrication constraints for
  3{D} us imaging,'' \emph{IEEE Transactions on Ultrasonics, Ferroelectrics,
  and Frequency Control}, vol.~64, no.~1, pp. 108--125, 2016.

\bibitem{roux20162}
E.~Roux, A.~Ramalli, P.~Tortoli, C.~Cachard, M.~C. Robini, and H.~Liebgott,
  ``2{D} ultrasound sparse arrays multidepth radiation optimization using
  simulated annealing and spiral-array inspired energy functions,'' \emph{IEEE
  Trans. Ultrason. Ferroelectr. Freq. Control}, vol.~63, no.~12, pp.
  2138--2149, 2016.

\bibitem{tekes2011optimizing}
C.~Tekes, M.~Karaman, and F.~L. Degertekin, ``Optimizing circular ring arrays
  for forward-looking {IVUS} imaging,'' \emph{IEEE Transactions on Ultrasonics,
  Ferroelectrics, and Frequency Control}, vol.~58, no.~12, pp. 2596--2607,
  2011.

\bibitem{roux2018experimental}
E.~Roux, F.~Varray, L.~Petrusca, C.~Cachard, P.~Tortoli, and H.~Liebgott,
  ``Experimental 3{D} ultrasound imaging with 2{D} sparse arrays using focused
  and diverging waves,'' \emph{Scientific Reports, Nature Publishing Group},
  vol.~8, no.~1, pp. 1--12, 2018.

\bibitem{cohen2018sparse}
R.~Cohen and Y.~C. Eldar, ``Sparse {D}oppler sensing based on nested arrays,''
  \emph{IEEE Transactions on Ultrasonics, Ferroelectrics, and Frequency
  Control}, vol.~65, no.~12, pp. 2349--2364, 2018.

\bibitem{cohen2018optimized}
------, ``Optimized sparse array design based on sum coarray,'' in
  \emph{International Conference on Acoustics, Speech and Signal
  Processing}.\hskip 1em plus 0.5em minus 0.4em\relax IEEE, 2018.

\bibitem{cohen2017sparse}
------, ``Sparse emission pattern in spectral blood doppler,'' in \emph{IEEE
  International Symposium on Biomedical Imaging}, 2017, pp. 907--910.

\bibitem{cohen2018coba}
------, ``Sparse convolutional beamforming for ultrasound imaging,'' \emph{IEEE
  Transactions on Ultrasonics, Ferroelectrics, and Frequency Control}, vol.~65,
  no.~12, pp. 2390--2406, 2018.

\bibitem{cohen2020sparse}
------, ``Sparse array design via fractal geometries,'' \emph{arXiv preprint
  arXiv:2001.01217}, 2020.

\bibitem{cohen2019sparse}
------, ``Sparse fractal array design with increased degrees of freedom,'' in
  \emph{IEEE International Conference on Acoustics, Speech and Signal
  Processing}, 2019, pp. 4195--4199.

\bibitem{puente1996fractal}
C.~Puente-Baliarda and R.~Pous, ``Fractal design of multiband and low side-lobe
  arrays,'' \emph{IEEE Transactions on Antennas and Propagation}, vol.~44,
  no.~5, p. 730, 1996.

\bibitem{werner1999fractal}
D.~H. Werner, R.~L. Haupt, and P.~L. Werner, ``Fractal antenna engineering: The
  theory and design of fractal antenna arrays,'' \emph{IEEE Antennas and
  Propagation Magazine}, vol.~41, no.~5, pp. 37--58, 1999.

\bibitem{werner2003overview}
D.~H. Werner and S.~Ganguly, ``An overview of fractal antenna engineering
  research,'' \emph{IEEE Antennas and Propagation Magazine}, vol.~45, no.~1,
  pp. 38--57, 2003.

\bibitem{feder2013fractals}
J.~Feder, \emph{Fractals}.\hskip 1em plus 0.5em minus 0.4em\relax Springer
  Science \& Business Media, 2013.

\bibitem{falconer2004fractal}
K.~Falconer, \emph{Fractal {G}eometry: {M}athematical {F}oundations and
  {A}pplications}.\hskip 1em plus 0.5em minus 0.4em\relax 2nd ed. Wiley, 2005.

\bibitem{jensen2002ultrasound}
J.~A. Jensen, ``Ultrasound imaging and its modeling,'' in \emph{Imaging of
  Complex Media with Acoustic and Seismic Waves}.\hskip 1em plus 0.5em minus
  0.4em\relax Springer, 2002, pp. 135--166.

\bibitem{jensen1999linear}
------, ``Linear description of ultrasound imaging systems,'' \emph{Notes for
  the International Summer School on Advanced Ultrasound Imaging, Technical
  University of Denmark July}, vol.~5, 1999.

\bibitem{steinberg1992digital}
B.~D. Steinberg, ``Digital beamforming in ultrasound,'' \emph{IEEE Transactions
  on Ultrasonics, Ferroelectrics, and Frequency Control}, vol.~39, no.~6, pp.
  716--721, 1992.

\bibitem{montaldo2009coherent}
G.~Montaldo, M.~Tanter, J.~Bercoff, N.~Benech, and M.~Fink, ``Coherent
  plane-wave compounding for very high frame rate ultrasonography and transient
  elastography,'' \emph{IEEE Transactions on Ultrasonics, Ferroelectrics, and
  Frequency Control}, vol.~56, no.~3, pp. 489--506, 2009.

\bibitem{roux20162d}
E.~Roux, ``2{D} sparse array optimization and operating strategy for real-time
  3{D} ultrasound imaging,'' Ph.D. dissertation, Lyon, 2016.

\bibitem{prager2010three}
R.~W. Prager, U.~Z. Ijaz, A.~Gee, and G.~M. Treece, ``Three-dimensional
  ultrasound imaging,'' \emph{Proceedings of the Institution of Mechanical
  Engineers, Part H: Journal of Engineering in Medicine}, vol. 224, no.~2, pp.
  193--223, 2010.

\bibitem{abo2004usefulness}
K.~Abo, T.~Hozumi, S.~Fukuda, Y.~Matsumura, M.~Matsui, K.~Fujioka, M.~Nakao,
  Y.~Takemoto, H.~Watanabe, T.~Muro \emph{et~al.}, ``Usefulness of
  transthoracic freehand three-dimensional echocardiography for the evaluation
  of mitral valve prolapse,'' \emph{Journal of Cardiology}, vol.~43, no.~1, pp.
  17--22, 2004.

\bibitem{kwan2014three}
J.~Kwan, ``Three-dimensional echocardiography: A new paradigm shift,''
  \emph{Journal of Echocardiography}, vol.~12, no.~1, pp. 1--11, 2014.

\bibitem{hoctor1990unifying}
R.~T. Hoctor and S.~A. Kassam, ``The unifying role of the coarray in aperture
  synthesis for coherent and incoherent imaging,'' \emph{Proceedings of the
  IEEE}, vol.~78, no.~4, pp. 735--752, 1990.

\bibitem{pal2010nested}
P.~Pal and P.~Vaidyanathan, ``Nested arrays: A novel approach to array
  processing with enhanced degrees of freedom,'' \emph{IEEE Transactions on
  Signal Processing}, vol.~58, no.~8, pp. 4167--4181, 2010.

\bibitem{lediju2011short}
M.~A. Lediju, G.~E. Trahey, B.~C. Byram, and J.~J. Dahl, ``Short-lag spatial
  coherence of backscattered echoes: Imaging characteristics,'' \emph{IEEE
  Transactions on Ultrasonics, Ferroelectrics, and Frequency Control}, vol.~58,
  no.~7, 2011.

\end{thebibliography}

\end{document}